\documentclass[onecolumn,showpacs,preprintnumbers,amsmath,amssymb]{revtex4}
\usepackage{graphicx,psfrag}
\usepackage{dcolumn}
\usepackage{bm}

\begin{document}

\title{Dispersive Light Propagation at Cosmological Distances: Matter Effects}

\author{David C. Latimer}

\affiliation{Department of Physics, University of Puget Sound,
Tacoma, WA 98416-1031 \footnote{Permanent Address}
}

\affiliation{
Department of Physics and Astronomy, University of Kentucky, 
Lexington, KY 40506-0055}

\newcommand*{\sech}{\mathop{\mathrm{sech}}\limits}
\newcommand*{\balpha}{\boldsymbol{\alpha}}
\newcommand*{\dilog}{\mathrm{Li}_2}
\newcommand{\qslash}{\not{\hbox{\kern-2.3pt $q$}}}
\newcommand{\kslash}{\not{\hbox{\kern-2.3pt $k$}}}
\newcommand{\pslash}{\not{\hbox{\kern-2.3pt $p$}}}
\newcommand{\gmn}{g^{\mu \nu}}
\newcommand{\Pslash}{\not{\hbox{\kern-2.3pt $P$}}}
\newcommand{\Pslashsup}{^\not{\hbox{\kern-0.5pt $^P$}}}

\begin{abstract}

Searches for dispersive effects in the propagation of light at cosmological 
distances have been touted as sensitive probes of Lorentz invariance violation (LIV) and
of theories of quantum gravity.  Frequency-dependent time lags between simultaneously emitted
pulses of light can signal a modification in the photon dispersion relation; however,
matter engenders the cosmos with a dispersive index of refraction  to similar
effect.   We construct a theoretical framework for the analysis of such 
effects, contrasting these dispersive terms with those from LIV models. 
We consider all matter, both luminous and dark.  
Though the only known mode of interaction for 
dark matter (DM) is gravitational, most models of dark matter also allow for electromagnetic interactions, if only at the one-loop level in perturbation theory.   Generically, the leading order dispersive effects due to matter scale with photon energy as $\omega^{-2}$ for a charged DM candidate and $\omega^2$ for a neutral DM candidate.  Terms linear in $\omega$ can arise in the index of refraction if parity and charge-parity asymmetries are present at the Lagrangian or system level.
Herein, we compute the index of refraction for a millicharged dark matter candidate at the one-loop level, 
a neutral scalar DM candidate introduced by Boehm and Fayet \cite{boehm1}, and 
the MSSM's neutralino.  For a neutral DM candidate, we determine that matter effects can compete with LIV effects that depend quadratically on energy whenever the photon energy is beyond $10^{29}$ GeV, well beyond the GZK cutoff.  The dispersive matter effects that scale linearly with $\omega$ are model dependent, and their existence results in circular birefringence. 
\end{abstract}

\maketitle

\section{Introduction}

Theories which violate Lorentz symmetry can result in a modified photon dispersion relation which manifests itself in an energy (or frequency) dependent speed of light in vacuum \cite{myers-pospelov,kosto,kost_mewes_em,kost_mewes_grb}.  At high energies, this vacuum speed of light could depend, to leading order, linearly or quadratically upon photon energy, and this energy dependence could either decrease or increase with energy.  Such modified dispersion relations are common in theories of quantum gravity \cite{gambini,Ellis:1999uh,alfaro,ellis_foam,ellis_string,Ellis:2004ay}.  As an experimental consequence, simultaneously emitted photons of different energies traveling over a long baseline will become temporally separated.   The energy dependence of the arrival time of photons from gamma ray bursts (GRBs) has been used  to place limits on the energy scale of quantum gravity or the scale at which possible LIV may occur \cite{AmelinoCamelia:1997gz,sch2,ellis,Boggs:2003kxa,Ellis:2005wr,jacobpiran,ellis2,Ellis:2009yx,AmelinoCamelia:2009pg,   Vasileiou:2013vra}.    GRBs are bright, violent bursts of high energy photons lasting on the order of fractions to hundreds of seconds, making them an apt source for study; indeed, their brightness makes observation at high redshift possible.   From GRB 080916C,  the observation of $\mathcal{O}$(GeV) photons probes LIV effects at the Planck scale \cite{fermi_nat}.

Though the matter density of space is low, it is not truly a vacuum.  Matter comprises nearly thirty percent of the universe's energy budget \cite{wmap9}, and over large scales, this matter density is relatively uniform \cite{wigglez}.  
The overwhelming majority of this matter is some yet unknown dark matter (DM).  Through its gravitational interaction, the existence of DM is established  from a concordance of astronomical observations from different cosmological epochs \cite{concord1,concord2,wmap9}.   Little is known about DM except for the fact that it is not relativistic, must be stable on the scale of billions of years, and interacts very little through non-gravitational channels.

Since the universe does contain matter, when taken {\em en masse}, this engenders the cosmos with an index of refraction.  Rather generically, this index of refraction is dispersive; that is, the phase velocity for light traveling through the medium is frequency dependent.  Similar to LIV effects, light traveling over long baselines through a dispersive medium will suffer a frequency-dependent time lag.  These matter effects could confound potential claims of LIV gleaned from GRB photon arrival times.  Herein, we assess the dispersive effects of the matter in the cosmos.  Since DM comprises the bulk of the matter in the universe, we will consider it in detail.

At the particle level, the index of refraction of a bulk medium arises from the forward scattering of photons on the particles that comprise the medium.  Photons directly couple to charged particles, yet the refractive index for neutral particles is also typically nonzero.  
  We emphasize that a dark-matter particle need not have an electric charge to 
scatter a photon.  For example, it can be electrically neutral but possess an electric or magnetic dipole moment, but such moments are not generic properties of DM candidates.  
Indeed, Majorana fermions cannot have permanent dipole moments, and the existence of a permanent electric dipole moment requires time-reversal and parity violation.
 For the  case in which the DM has no permanent dipole moments, the DM medium can still have a nonzero polarizability
and, hence, index of refraction, provided that the DM particle couples to other electromagnetically charged particles.  Classically, the polarizability of a substance is a measure of an external electric field's ability to  induce dipole moments in the molecules that comprise the substance.  In the simplest case, say for a spherically symmetric atom, an external electric field exerts opposite forces on the protons and electrons, resulting in a nonzero dipole moment parallel to the external field.  At leading order, this substance's polarizability is proportional to the applied electric field; however, higher order contributions depend quadratically, etc., on the field strength.  If a structureless neutral particle couples to electrically charged particles, then an electric field can induce an analogous effective dipole moment, despite the fact that the neutral particle is not composed of charged constituents.
Many DM candidates do couple to charged particles.  In fact, some indirect searches for DM involve the detection of high energy photons produced in a DM-DM annihilation event.  
  Since the Compton amplitude can be related by crossing symmetry 
to the amplitude for dark-matter annihilation into two photons, then any dark-matter model which gives rise to an indirect detection signal via $\chi \bar \chi \to \gamma\gamma$~\cite{Ullio:2002pj} can also drive the index of refraction of light from unity.

For a given DM model, one must compute the forward Compton amplitude to determine the particle's contribution to the refractive index; however, in a low photon-energy expansion, some model independent statements can be made about the amplitude. 
If  charge conjugation, parity, and time-reversal symmetries are conserved, then the leading order contributions to the forward scattering amplitude are model independent, attributable to the charge, mass, and magnetic dipole moment of the scatterer \cite{GGT,low,GG,G,lapidus}.  Should these discrete symmetries be violated, then additional contributions to the forward scattering amplitude arise. By including parity- and time-reversal violating terms in the Hamiltonian,  a more general form for the scattering amplitude has been explored in Ref.~\cite{bedaque,chen,gorchtein}.  In general, these new terms are model dependent, and it turns out that only parity violating interactions contribute to the amplitude in the forward limit.   Such parity-violating terms can account for permanent electric dipole moments and anapole interactions.  

 In Ref.~\cite{millicharge},  we considered the leading order behavior of the forward Compton amplitude and used limits from the non-observation of dispersion attributable to charged matter to place constraints on the charge to mass ratio of sub-eV dark matter particles.   Herein, we consider higher order contributions to the index of refraction.  In particular, we  compute higher order contributions to the index from millicharged matter, namely its quasi-static polarizability -- so called because a charged particle's polarizability diverges in the (static) limit of vanishing frequency \cite{holstein_sum}.  
 We  also consider two models of neutral dark matter--the Boehm-Fayet (BF) scalar DM model \cite{boehm1} and the neutralino of the Minimal Supersymmetric Standard Model (MSSM).  We  compute the full forward Compton amplitude to first order in perturbation theory for both DM candidates; this will result in the polarizability of the DM medium.   At low photon energies, we find that the explicit computations yield the form expected from purely analytic considerations.

The electromagnetic form factors of DM have been considered elsewhere.  Stringent bounds on the fractional charge of more massive (MeV to GeV) dark matter candidates have been established by requiring that the charge not be so large as to affect the density perturbations at recombination from the CMB spectrum \cite{mcdermott}.    Additionally, bounds on the electric and magnetic dipole moments have been established through both theoretical and experimental considerations including DM relic abundance, non-observation of DM-nuclei interactions in direct detection experiments,  precision measurements,  and non-production at colliders \cite{Sigurdson:2004zp,heo_mdm,heo_edm,barger_dm,barger_lhc}.  Others have determined bounds on the polarizability of DM which sets the scale for the refractive index in the static limit, i.e., for low photon energies.  In Ref.~\cite{pospelov}, the authors used direct detection searches to limit the polarizability of weakly interacting massive particles with masses on the order of 100 GeV.  Additionally, in Ref.~\cite{Cotta:2012nj}, the authors take an effective-field-theory approach to constrain the interaction between dark matter and the electroweak gauge bosons.  Their constraints on the effective coupling for a photon-DM contact term can be mapped to the polarizability of the DM particle.  As for these polarizability constraints, the leading order term only sets the scale for the refractive index's deviation from unity because it is independent of omega. 
To assess, the dispersive nature of the DM medium, higher order contributions to the polarizability are needed, which we compute below.

For neutral matter, dispersion rather generically depends quadratically upon photon energy.
Because of collider constraints on the existence of heavy of charged, exotic particles, the dispersion due to neutral DM is extremely small, and GRB photon arrival time currently provides no meaningful constraints on DM models.  
 The  redshift dependence for time lags due to a DM medium is distinct from those due to LIV effects, yet it is possible that matter effects could confound claims of the existence of a modified photon dispersion relation that is quadratic in energy.  Herein, we determine the energy scale at which matter effects become comparable in size with such LIV effects.  As for dispersive matter effects which scale linearly with energy, we find that these arise if there are parity violating terms in the forward scattering amplitude.  These terms can produce a circularly birefringent medium; that is, the index of refraction is different for right- and left-handed polarized light.  These effects could potentially be confounded with a modified photon dispersion relation which scales linearly with energy, but based upon one model dependent calculation, these matters effects appear to be small.

\section{Observational framework}

The relationship between the index of refraction $n_\lambda (\omega)$ and the 
coherent forward scattering amplitude $f_{\lambda \lambda}(0)$ for light of frequency $\omega$ 
and polarization $\lambda$ is given by the Lorentz relation~\cite{goldwatson}
\begin{equation}
n_\lambda (\omega)  = 1 + \frac{2\pi N}{\omega^2} f_{\lambda \lambda}(0) \label{qm_n},
\end{equation} 
where $N$ denotes the number density of scatterers, and we work in units of
$c = \hbar =1$ throughout. The relationship can be derived by considering
the transmission of a wave through a thin slab of material 
at rest~\cite{goldwatson,fermi}, where the deviation of $n$ from unity arises from the interference between the incident and scattered waves. 
To relate the completely coherent quantum mechanical forward scattering amplitude, $f_{\lambda \lambda}(0)$,
to the analog in quantum field theory, ${\cal M}_{\lambda \lambda}(k,p \to k,p)$, we compute 
the differential cross section for a generic forward scattering event in the scatterer's rest frame 
and match it with  $\mathrm{d} \sigma/ \mathrm{d} \Omega = \vert f_{\lambda \lambda}(0) \vert^2$. 
Using the conventions of Ref.~\cite{ps}, we determine 
$f_{\lambda \lambda}(0) = {\cal M}_{\lambda \lambda} (k,p \to k,p)/8\pi m$, 
where the overall phase is fixed by demanding that the field theoretic amplitude obey the optical theorem.  
We thus have 
\begin{equation}
n_\lambda (\omega)  = 1 + \frac{ \rho}{4m^2 \omega^2} \mathcal{M}_{\lambda \lambda}(k, p \to k,p) ,
\label{qft_n}
\end{equation}
with  $\rho=mN$ the mass density of the scatterers. Note that the 
amplitude is evaluated in the dark-matter rest frame 
so that $p=(m,\mathbf{0})$ and $k=(\omega, \omega \hat{\mathbf{n}})$ where $\hat{\mathbf{n}}$ points in the direction of   light propagation.

To determine the index of refraction for a specific dark matter model, one must compute the forward Compton amplitude for that model; however, there are significant model-independent features of the index of refraction that we wish to highlight.
Assuming Lorentz invariance and  conservation of the discrete symmetries $C$, $P$, and $T$, we can factor the forward Compton amplitude as follows~\cite{GGT,low,GG,G,lapidus} 
\begin{equation} 
\mathcal{M}_{\mu \lambda}(k, p \to k,p) = g (\omega)\,
\boldsymbol{\epsilon}_\mu^* \cdot \boldsymbol{\epsilon}_\lambda
+ i h (\omega)\, \boldsymbol{\mathcal{S}} \cdot (\boldsymbol{\epsilon}_\mu^* \times \boldsymbol{\epsilon}_\lambda) , \label{for_amp}
\end{equation}
where ${\boldsymbol{\mathcal S}}$ is the spin operator associated with the dark-matter particle
and $\boldsymbol{\epsilon}_\lambda$ ($\boldsymbol{\epsilon}_\mu$) is the polarization vector associated with
the photon in its initial (final) state. We note that the photon is transverse, so that $\boldsymbol{\epsilon}_\lambda \cdot \hat{\mathbf{n}}= \boldsymbol{\epsilon}_\mu \cdot \hat{\mathbf{n}} =0$.  If these discrete symmetries are violated, then additional terms may appear in the forward Compton amplitude.  In Ref.~\cite{gorchtein}, we find the general structure of the Compton amplitude for processes which violate time reversal symmetry.    For real, transverse photons, $T$-violating processes do not contribute to the amplitude in the forward limit.   On the other hand, parity violating interactions do contribute.  From Refs.~\cite{bedaque,chen}, we find
\begin{equation}
 \mathcal{M}\Pslashsup_{\mu \lambda}(k, p \to k,p) =  g\Pslashsup (\omega)\,  (\boldsymbol{\mathcal{S}} \cdot  \hat{\mathbf{n}})\,
(\boldsymbol{\epsilon}_\mu^* \cdot \boldsymbol{\epsilon}_\lambda)
+ i h\Pslashsup (\omega)\, \hat{\mathbf{n}} \cdot (\boldsymbol{\epsilon}_\mu^* \times \boldsymbol{\epsilon}_\lambda). \label{pviol_amp}
\end{equation}

At leading order in a perturbative calculation of the forward Compton amplitude, there exists a nonzero threshold photon energy $\omega_\text{th} >0$ at which virtual particles in the scattering process go on mass shell.  Below this inelastic threshold, the functions $g(\omega)$ and $h(\omega)$ are shown to be both real and even and odd, respectively, under the map $\omega \mapsto -\omega$.  Additionally, in a power series expansion of these functions about $\omega$, it is shown that their leading order behavior is determined by the static properties of the scatterer, namely its charge, mass, and magnetic moment.  In particular, we have
\begin{eqnarray}
g(\omega)  &=& - 2\varepsilon^2 e^2 + \mathcal{O}(\omega^2) ,\\
h(\omega)  &=&  - 2m \left(\frac{\mu}{\mathcal{S}} - \frac{\varepsilon e}{m}\right)^2 \omega + \mathcal{O}(\omega^3) ,
\end{eqnarray}
where $\varepsilon$ is the charge of the scatterer (in units of $e$) and $\mu$ is its magnetic moment.  
This low energy behavior holds even if the scatterers are loosely bound composite particles \cite{brodsky}. 
For the $P$-violating terms in the amplitude, the leading order behavior of the functions  $g\Pslashsup (\omega)$ and $h\Pslashsup (\omega)$ are model dependent.   As an example, for photon-nucleon scattering, the leading order behavior for these functions, in the center-of-mass frame, is \cite{chen}
\begin{eqnarray}
g\Pslashsup (\omega) &\sim&- \frac{e^2 g_A h^{(1)}_{\pi NN} M_N \omega^2}{ 3 \sqrt{2} \pi^2 F_\pi  m_\pi^2} , \label{gslash_lot} \\
h\Pslashsup (\omega) &\sim& \frac{e^2 g_A h^{(1)}_{\pi NN} \mu_n \omega^3}{ 6\sqrt{2} \pi^2 F_\pi  m_\pi^2}. \label{hslash_lot}
\end{eqnarray}
Here, $g_A$ is the nucleon-pion coupling constant; $h^{(1)}_{\pi NN}$  is the $P$-violating isovector pion-nucleon coupling; $M_N$ is the nucleon mass; $F_\pi$ is the pion decay constant; and $\mu_n$ is the neutron magnetic moment in units of nuclear magnetons.  In an expansion about $\omega$, the leading order contribution to the scattering amplitude takes the same form in both the center-of-mass frame and the scatterer's rest frame (though this is {\em not} true for higher order terms in the expansion).  This low-energy expansion is valid for photon energies well below the inelastic pion threshold, $\omega \ll 140$ MeV.

We note that the terms in the forward amplitude, Eqs.~(\ref{for_amp},\ref{pviol_amp}), which depend on the cross product of the polarization vectors, $\boldsymbol{\epsilon}_\mu^* \times \boldsymbol{\epsilon}_\lambda$, change the polarization of the incident photon.  If the incident electromagnetic wave is linearly polarized, then scattered waves of this ilk cannot completely coherently interfere with the incident wave, which is why, per Eq.~(\ref{qft_n}), these terms do not contribute to the index of refraction.  However, we can extend our considerations to include optically active media.   
In this case, the $h(\omega)$ and $h\Pslashsup(w)$ terms result in birefringence for circularly polarized light; that is, the index of refraction, $n_+$, for postive-helicity photons differs from that, $n_-$, for negative-helicity photons.  Let us suppose that the electromagnetic wave is traveling in the direction $\hat{\mathbf{n}}= \hat{\mathbf{e}}_3$; then, it can have linear polarization $ \hat{\mathbf{e}}_1$ or $ \hat{\mathbf{e}}_2$.  But, we could equally as well consider states of circular polarization $\boldsymbol{\epsilon}_\pm =\frac{1}{2}(\hat{\mathbf{e}}_1 \pm i \hat{\mathbf{e}}_2) $.  In this basis, the $P$-violating forward Compton amplitude is 
\begin{equation}
 \mathcal{M}\Pslashsup_{\pm \pm}(k, p \to k,p) =  g\Pslashsup (\omega)\,  \boldsymbol{\mathcal{S}} \cdot  \hat{\mathbf{n}}
\pm h\Pslashsup (\omega), \label{pviol_amp_circ}
\end{equation}
with $\mathcal{M}\Pslashsup_{\pm \mp}(k, p \to k,p)=0$.    Hence, the phase speed of the photons differs according to helicity.

For the moment, we focus upon the amplitude which conserves the discrete symmetries, Eq.~(\ref{for_amp}).
As for $g(\omega)$, by imposing the additional assumption of causality
and using the analyticity and unitarity of the scattering matrix, dispersion relations emerge for $g(\omega)$ 
and $h(\omega)$ \cite{GGT,G}
\begin{eqnarray}
\mathrm{Re} \, g(\omega) - \mathrm{Re}\, g(0)
 &=& \frac{4m\omega^2}{\pi} \int_0^\infty 
\mathrm{d}\omega'
\frac{\sigma(\omega')}{{\omega'}^2 - \omega^2} \label{disp_rel_g}
 ,\\
\mathrm{Re}\, h(\omega) - \omega \mathrm{Re}\, h'(0) &=& \frac{4m \omega^3}{\pi} \int_0^\infty 
\mathrm{d}\omega'
\frac{\Delta \sigma(\omega')}{\omega'({\omega'}^2 - \omega^2)} \,,
\label{disp_rel_h}
\end{eqnarray}
where we use the optical theorem to replace $\mathrm{Im}\, g(\omega)$ with 
the unpolarized cross section $\sigma$.  We define $\Delta  \sigma \equiv \frac{1}{2}( \sigma_p - \sigma_a)$.  
where $\sigma_p$ ($\sigma_a$) is 
the cross section for parallel (antiparallel) photon and scatterer spins.  The low energy theorem sets $g(0) = -2 \varepsilon^2 e^2$ and $h'(0) = - 2m \left(\mu/\mathcal{S} - \varepsilon e/m \right)^2$.
Recall, at leading
order in perturbation theory, $g(\omega)$ and $h(\omega)$ are real for $\omega < \omega_\text{th}$, so that
the lower limit of this integral is merely formal, since $\sigma(\omega) \equiv 0$ below the inelastic
threshold (at the same order in perturbation theory).  
As such, the integral actually commences at $\omega_{\rm th}$ so that it is well-posed at $\omega'=0$. 
The  relationship for $g(\omega)$ is the quantum field theoretic analogue 
of the Kramers-Kronig~\cite{kramers_kronig} relationship of optics.

In the limit $\omega \ll \omega_{\rm th}$, we can expand $g(\omega)$, Eq.~(\ref{disp_rel_g}),  as a series of even powers of $\omega$ with {\em positive} coefficients  ${\cal M}_{\lambda \lambda} = \sum_{j=0} A_{2j} \omega^{2j}$, we thus have 
\begin{equation}
n(\omega) = 1 + \frac{ \rho}{4m^2 \omega^2}\left( A_0  + A_2\omega^2 + \dots \right)\,,
\label{n_expans}
\end{equation}
where $A_0 = - 2\varepsilon^2 e^2$ and $A_{2j}\ge 0$. The terms in ${\cal O}(\omega^2)$ and higher are associated
with the polarizability of the dark-matter candidate.  This is the expected refractive index for a medium which is not optically active.   Again, for $\omega \ll \omega_\text{th}$, we can expand $h(\omega)$, Eq.~(\ref{disp_rel_h}),  as a power series in $\omega$.  Below threshold, this yields a series of odd powers in $\omega$.  Inserting this into the index formula, we see that we can achieve a birefringent term that scales as $\omega^{-1}$; this term is associated with the scatterer's anomalous magnetic dipole moment.  Higher order terms can result in birefringence linear $\omega$.  Additionally, the leading term in the parity-violating contribution from nucleons, Eq.~(\ref{hslash_lot}), yields an odd power of $\omega$ in the refractive index; in this case, the dependence is linear. 
Thus, we see that in optically active media there can be terms in the index of refraction which are linear in $\omega$,
similar to some quantum gravity models. 

The assumptions that lead to the model-independent form of the index of refraction in Eq.~(\ref{n_expans}) are actually too restrictive for our purposes as even a Standard Model background of particles is not symmetric under the discrete symmetries.  As an example, consider a background of neutrinos and antineutrinos.  At the Lagrangian level of this system, we can, to a good approximation, take $CP$ as conserved; however, given the chiral nature of the weak interactions, parity is maximally violated.  If the number density of neutrinos differs from that of the antineutrinos in the background, then this medium will be birefringent and odd powers of $\omega$ will appear in the index of refraction \cite{nieves_pal,abb_rep1,abb_rep2}.   Such a background is asymmetric under a $CP$ transformation.  If this asymmetry is removed by equalizing the neutrino and antineutrino number density, then the medium is no longer optically active and, below threshold, only even powers of $\omega$ are present in the refractive index, consistent with Eq.~(\ref{n_expans}).  
In general, it was shown in Ref.~\cite{nieves_pal} that both $P$ {\em and} $CP$ asymmetry must be present in either the Lagrangian or background (or both) in order for the index of refraction to be birefringent with odd powers of $\omega$; failing that, Eq.~(\ref{n_expans}) will hold.   From a SM standpoint, it is easy to conceive that the background matter is $CP$ asymmetric -- after all we live in a universe predominantly comprised of {\em matter}, not {\em antimatter}.  In order to achieve an optically active medium, we merely need either $P$ violating processes in the light-matter interactions as in Eq.~(\ref{pviol_amp}) or a mechanism to make the background $P$ asymmetric.  An example of such a background would be a polarized medium which could occur, for instance, in a region containing a background magnetic field \cite{GM_faraday}.  In that case, one would expect leading order contribution to $h(\omega)$, attributable to the particle's magnetic dipole moment, to induce birefringence. 

We now turn to the manner in which 
we can realize observational constraints on the cosmic index of refraction.   The refractive index generated by the completely coherent portion of the forward Compton amplitude is generic to all matter, so we will confine our attention, for the moment, with the observation procedure one may use to  realized constraints on the coefficients $A_{2j}$.
We will return to a discussion  of an optically active medium later.
The index of refraction fixes the phase velocity $v_p$ in the medium; it is 
$v_p=\omega/k=1/\tilde n$ with $\tilde n \equiv \hbox{Re}\,n$. 
Observable dispersive effects in light propagation are controlled, rather, by the group velocity
$v_g$, for which $v_g = \mathrm{d}\omega/\mathrm{d}k= 
(\tilde n + \omega (\mathrm{d}\tilde n/\mathrm{d} \omega))^{-1}$. Thus the light emitted 
from a source 
at a particular time at a distance $\ell$ away possesses a 
frequency-dependent arrival 
time $t(\omega)$, namely 
$t(\omega)= \ell(\tilde n + \omega \mathrm{d}\tilde n/\mathrm{d}\omega)$, or 
\begin{equation}
\!\!\!t(\omega) = \ell \left( 1 + \frac{\rho}{4 m^2} \left( \frac{-A_0}{\omega^{2}} + 
A_2 + 3 A_4 \omega^2 + 
 {\cal O}(\omega^4) \right) \right) \,. 
\label{tlag}
\end{equation}
For sources at cosmological distances, 
we must account for the impact of an expanding universe on the arrival time~\cite{jacobpiran}. 
As we look back to a light source at redshift $z$, we note that the 
dark-matter density accrues a scale factor of $(1+z)^3$, whereas 
the photon energy is blue shifted by a factor of $1+z$ relative to its present-day 
value $\omega_0$~\cite{jacobpiran}. Thus we have the arrival time $t(\omega_0,z)$ for
light of observed angular 
frequency $\omega_0$ from a source with red shift $z$: 
\begin{equation}
\!t(\omega_0, z) = \!\int_0^z \!\frac{\mathrm{d}z'}{H(z')} 
\left[ 1 + \frac{\rho_0(1+z')^3}{4 m^2} \left( \frac{-A_0}{ (1+z')^2\omega_0^{2}}  
+ A_2 + 3 A_4 (1 + z')^2 \omega_0^2 + 
{\cal O}(\omega_0^4) \right) \!\right] 
\label{cosmictlag}
\end{equation}
with the Hubble rate $H(z') = H_0 \sqrt{(1+z')^3\Omega_M + \Omega_\Lambda}$. 
The cosmological parameters determined through the combined analysis of WMAP nine-year data 
in the $\Lambda$CDM model with  distance measurements from Type Ia supernovae (SN) and with baryon acoustic oscillation
information from the distribution of galaxies can be found in Ref.~\cite{wmap9}. The Hubble constant
today is $H_0 = 69.32 \pm 0.80 \,\hbox{km/s/Mpc}$, whereas 
the fraction of the energy density
in matter relative to the critical density today is $\Omega_M = 0.2865^{+0.0097}_{-0.0096}$ and 
the corresponding fraction of the energy density 
in the cosmological constant $\Lambda$ is 
$\Omega_\Lambda = 0.7135^{+0.0095}_{-0.0096}$~\cite{wmap9}. 
Not only do distant astrophysical sources provide an increased baseline over which frequency-dependent time lags can develop; the photon frequency and dark matter density's dependence upon redshift provide an additional cosmological lever arm.  For the coefficient $A_{2j}$ in Eq.~(\ref{cosmictlag}), the integral over redshift is given by
\begin{equation}
K_{2j}(z)= \int_0^z \mathrm{d}z' \frac{(1+z')^{2j+1}}{H(z')}
\end{equation} 
so that higher order $j$ have more factors of $1+z$ in the numerator.

Various strategies must be employed to isolate the coefficients 
$A_{2j}$ in Eq.~(\ref{cosmictlag}). 
To start we note that the $A_2$ term has no frequency dependence, 
so that its effects are unobservable with our method, though its consequences
have been explored in Ref.~\cite{sl_nu}. 
The remaining terms can be constrained by dispersive effects. 
The term that scales as $\omega_0^{-2}$ is best constrained through the arrival time difference 
of a gamma-ray pulse from a GRB and its radio afterglow \cite{millicharge}. 
The terms in positive
powers of $\omega_0$ are best constrained through arrival time differences between 
optical and gamma-ray
pulses of varying energy. 

\section{Illustrative models}

Though the observational program can be carried out independent of a particular dark matter model, we compute the forward Compton amplitude for three explicit particles:   a millicharged fermion, the BF scalar DM particle, and MSSM's neutralino.  These explicit models will demonstrate that the generic analytic behavior of the refractive index holds.

\subsection{Charged matter}
\begin{figure}[th]
\includegraphics[width=3in]{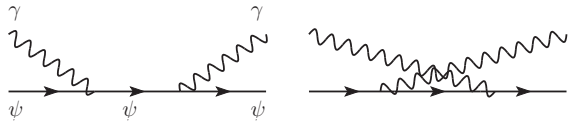}
\caption{Feynman diagrams for the $\mathcal{O}(\hat \alpha)$ contribution to the forward Compton amplitude for a charged fermion $\psi$.   The second diagram is the ``crossed-photon" version of the first diagram.
\label{fig1}}
\end{figure}

Millicharged particles arise naturally in models where the SM photon kinetically mixes with a new massless gauge boson \cite{mirror}.  As discussed earlier, there are limits on 
heavy millicharged DM \cite{mcdermott}.  A summary on the existence of particles with fractional electric charge can be found in Ref.~\cite{davidson}.
In Ref.~\cite{millicharge}, we constrain the electric charge of light dark matter through the non-observation of dispersion due to charged matter.  For matter of charge  $\varepsilon e$ and mass $m$, the leading order contribution to the forward Compton amplitude is set by the low energy theorem to be $A_0=-2 \varepsilon^2 e^2 = -8 \pi \hat{\alpha}$ where we define $\hat\alpha = \varepsilon^2 e^2/4\pi$.    This is the result upon evaluating the Feynman diagrams for forward scattering on a charged fermion, Fig.~\ref{fig1}.   In this figure, the second diagram is the ``crossed-photon" version of the first diagram; we will omit the ``crossed" diagrams in the figures that follow.

Higher order contributions to the amplitude will  yield terms which are dependent upon the photon energy.  Such contributions are model dependent, requiring input from the structure of the scatterer and its coupling to other particles.  We make no assumptions of structure or couplings and only calculate the na\"{i}ve $\mathcal{O}(\hat \alpha^2)$ contribution to the forward scattering amplitude, $\mathcal{M}^{(2)}$.   A word of caution is in order because the low energy expansion of the forward amplitude, $\mathcal{M}= \sum_{j=0} A_{2j} \omega^{2j}$, is not relevant at $\mathcal{O}(\hat \alpha^2)$ for a structureless charged particle.  This is because the inelastic threshold for a photon scattering  on a charge particle commences at $\omega=0$ at this order in perturbation theory \cite{llanta}.  This follows from a simple application of the optical theorem.  The optical theorem relates the total cross section for a scattering process to the imaginary part of the forward amplitude
$ \mathrm{Im}\, \mathcal{M}(k,p \to k,p) = E_\text{cm} p_\text{cm} \sigma_\text{tot} $ where $E_\text{cm}$ is the energy in the center of momentum frame and $p_\text{cm}$ is the momentum of one the particles in this frame.
 At low photon energies, the Thomson cross section is valid for this scattering process $\sigma_\text{tot} = 8 \pi \hat\alpha^2/3m^2$.    Using this cross section, we would expect that the imaginary part of the forward amplitude, at low photon energies, to be $ \mathrm{Im}\, \mathcal{M}(k,p \to k,p) = 16 \pi \hat\alpha^2\omega/3m$.
Since $\mathrm{Im}\, \mathcal{M}^{(2)}$ is proportional to $\omega$ in the limit of vanishing photon energy, then it is clear that the inelastic threshold for photon scattering  is $\omega=0$ at this order in perturbation theory.

\begin{figure}[ht]
\includegraphics[width=3in]{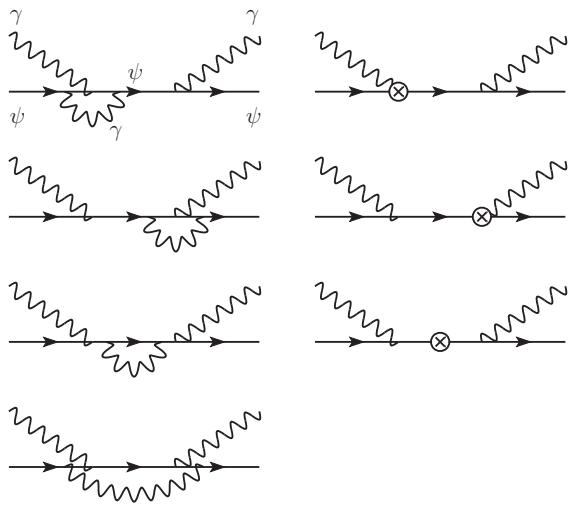}
\caption{Feynman diagrams for the $\mathcal{O}(\hat \alpha^2)$ contribution to the forward Compton amplitude for a charged fermion $\psi$.   The graphs with ``crossed" photons are not shown. 
\label{fig2}}
\end{figure}

In Fig.~\ref{fig2}, we present the Feynman graphs needed to determine $\mathcal{M}^{(2)}$; note, the graphs with ``crossed" photons are not shown.  The infrared divergences generated by the internal photon lines are regulated by giving the photon a small fictitious mass $\mu$; however, upon summing all the diagrams, this photon mass vanishes in the forward limit so that the physical process is independent of $\mu$.  The ultraviolet divergences generated by the vertex and propagator corrections are regulated, upon renormalization, by the vertex and propagator counterterms.  We use FORM \cite{form} to aid our algebraic manipulations.

To set the notation, let the fermion four momentum be $p$ with $p^2=m^2$ and the photon four momentum be $k$ with $k^2=0$. 
The real part of the $\mathcal{M}^{(2)}$ contribution to the forward Compton amplitude is
\begin{eqnarray}
\mathrm{Re}\, \mathcal{M}^{(2)} &=&   \hat \alpha^2  \Bigg[    6 + \frac{m^2}{s} +  \frac{m^2}{u}+ \log\left ( \frac{2 k \cdot p}{m^2}\right) \Bigg(  \frac{m^4}{s^2} +  \frac{m^4}{u^2} -2 \Bigg) + 2 \Bigg( \dilog\left(\frac{u}{m^2}\right) + \dilog\left(\frac{s}{m^2}\right)
          - 2\dilog(1) \Bigg) \nonumber \\
 &&      + \frac{4 m^2}{k\cdot p}   \Bigg( \dilog\left(\frac{u}{m^2}\right) - \dilog\left(\frac{s}{m^2}\right) \Bigg)-  \frac{4 m^4}{(k \cdot p)^2}  \Bigg(   \dilog\left(\frac{u}{m^2}\right) + \dilog\left(\frac{s}{m^2}\right) - 2 \dilog(1)  \Bigg) \Bigg] 
\label{rech}
\end{eqnarray}
with the usual Mandelstam variables $s=(p+k)^2$ and $u=(p-k)^2$.  
We employ the definition of the dilogarithm found in Ref.~\cite{lewin}
\begin{equation}
\dilog(z) = - \int_0^z \mathrm{d} z' \frac{\log(1-z')}{z'}.
\end{equation}

For completeness, we include the imaginary part of the amplitude
\begin{equation}
\mathrm{Im}\, \mathcal{M}^{(2)} =  \hat \alpha^2   \pi \Bigg[
       1+ 6 \frac{m^2}{s} -  \frac{m^4}{s^2} + 5 \frac{m^2}{k \cdot p} + 3 \frac{m^4}{(k\cdot p)   s} + 2 \log \Big( \frac{s}{m^2} \Big) \Bigg(1 - 2\frac{m^2}{k \cdot p} - 2 \frac{m^4}{(k \cdot p)^2} \Bigg) \Bigg].
\end{equation}  
As a check, we reproduce the Thomson cross section via the optical theorem.
Working in the scatterer's rest frame with $k=(\omega,\omega \mathbf{\hat{n}})$ and $p=(m, \mathbf{0})$, we find that, in the limit of vanishing photon energy $\omega \to 0$, the imaginary part of the amplitude behaves as $\mathrm{Im}\, \mathcal{M}^{(2)} \to 16\pi \hat \alpha^2  \omega/3m + \mathcal{O}(\omega^3)$, which agrees with our expectations.

Returning to the real part of the amplitude, we will focus upon the leading order contributions to the polarizability of the fermion; in particular,  in an expansion about $\omega$, we find
\begin{equation}
\mathrm{Re}\, \mathcal{M}^{(2)} \approx  \hat \alpha^2  \Bigg[  \frac{\omega^2}{m^2}  \Bigg( \frac{44}{9} + \frac{64}{3} \log\left ( \frac{2 \omega}{m}\right) \Bigg)  \Bigg].
\label{quasipol}
\end{equation}
The expression in Eq.~(\ref{quasipol}) which is proportional to $\omega^2$ can only define a quasi-static polarizability due to the presence of the $\log(\omega)$ term which leads to a divergence in the static limit \cite{holstein_sum}; we note that our expression for the quasi-static polarizability extracted from Eq.~(\ref{rech}) is equivalent to the analogous expression in Refs.~\cite{llanta,holstein_sum}.

Calculating the local group speed for light traveling in the charged medium, we have to $\mathcal{O}(\hat \alpha^2)$
\begin{equation}
v_g^{-1}=1+\frac{\rho}{4m^2}\mathrm{Re}\,\left( \frac{8\pi\hat \alpha}{\omega^2}+ \frac{1}{\omega}\frac{\mathrm{d}\mathcal{M}^{(2)}}{\mathrm{d}\omega} - \frac{\mathcal{M}^{(2)} }{\omega^2}\right).  \label{vg_correction}
\end{equation}
To assess the effect of the correction upon the group velocity, we estimate its size separately for two regimes, small and large $\omega$.  For $\omega \ll m$, one has 
\begin{equation}
 \frac{1}{\omega}\frac{\mathrm{d}\mathcal{M}^{(2)}}{\mathrm{d}\omega} - \frac{\mathcal{M}^{(2)}}{\omega^2} \approx \hat \alpha^2 \Bigg[ \frac{1}{m^2}  \Bigg( \frac{236}{9} + \frac{64}{3} \log\left ( \frac{2 \omega}{m}\right) \Bigg) \Bigg].
\end{equation}
For small $\omega$, we see that a logarithmic term modifies the usual dispersion of light in the medium.  The variation in the logarithm is rather slow relative to the leading $\mathcal{O}(\omega^{-2})$ dispersion; furthermore, the logarithm term is suppressed by the additional factor $ \hat \alpha \omega^2/m^2$ relative to the leading order term in the group velocity. This correction will have little impact upon the arrival time of radio frequency waves from a distant GRB. 
Recalling the analysis from Ref.~\cite{millicharge}, we sought a frequency dependent time lag between the arrival of high energy photons from a GRB and its radio afterglow.  From the non-observation of such a time lag (consistent with millicharged dark matter), we  limit the charge-to-mass ratio of dark matter to be $| \varepsilon | /m < 10^{-5}$ eV$^{-1}$.   To assess the size of the correction to the leading order dispersive factor, consider a 1 GHz radio wave which corresponds to a photon energy of $\omega \sim 4 \times 10^{-6}$ eV; then, the next term is negligible, representing a percent correction of, at most,  $\hat \alpha \omega^2/m^2 \sim 10^{-23}$ multiplied by the logarithm.

We now consider the other extreme of large photon energies, namely $\omega \gg m$.  In this limit, the Mandelstam variables behave as $s\to 2\omega m$ and $u \to -2\omega m$.
We  recall the behavior of the dilogarithms from Ref.~\cite{lewin}.  For $x\gg 2$, one has the limit
\begin{equation}
\mathrm{Re}\, \dilog(x)
\to  \frac{\pi^2}{3}-\frac{1}{2}\log^2(x), \label{diloglim1}
\end{equation}
and for $x \ll -1$, one finds
\begin{equation}
\dilog(x) \to  -\frac{\pi^2}{6}-\frac{1}{2}\log^2\vert x \vert.  \label{diloglim2}
\end{equation}
Using Eqs.~(\ref{diloglim1},\ref{diloglim2}) in the amplitude Eq.~(\ref{rech}), the correction to the group velocity in the high energy limit behaves as
\begin{equation}
 \frac{1}{\omega}\frac{\mathrm{d}\mathcal{M}^{(2)}}{\mathrm{d}\omega} - \frac{\mathcal{M}^{(2)}}{\omega^2} \approx  \frac{\hat \alpha^2}{\omega^2}  \Bigg[- 8+ \frac{\pi^2}{3}  + 2\log\left(\frac{2\omega}{m}\right) + 2\log^2\left(\frac{2\omega}{m}\right)\Bigg].  \label{hi_E_charged}
\end{equation}
For sufficiently large $\omega$, this correction term will dominate the leading order term in the dispersion relation.  Crudely, we may estimate that for photon energies $\omega \gtrsim m/2 \exp[\sqrt{4\pi/ \hat\alpha}]$ the correction $\omega^{-2} \log^2 (2 \omega/m)$ will be the dominant source of dispersion.  Again, from Ref.~\cite{millicharge}, we limit the charge-to-mass ratio of dark matter to be $| \varepsilon | /m < 10^{-5}$ eV$^{-1}$.  To assess the smallest value of $\omega$ where the dispersive corrections will dominate, we consider the unrealistic possibility of a particle with unit electric charge $\varepsilon=1$, requiring $m > 10^5$ eV.  Then, the correction to the $\omega^{-2}$ dispersion dominates for a photon energy of $\omega \gtrsim 5 \times 10^{22}$ eV; this is a prohibitively high energy, orders of magnitude greater than the GZK cutoff \cite{gzk}.   Alternatively, we consider the size of the correction for more reasonable photon energies, $\omega \sim 1$ TeV.  For $\varepsilon = 10^{-5}$ and $m= 1$ eV, we find the correction to the usual $\omega^{-2}$ behavior to be on the order of a few parts in $10^{11}$.  As such, for millicharged DM, these higher order corrections are not appreciable for the photon energies
under consideration.

This higher order process in perturbation theory is na\"ive in that we assume a theory with only a charged fermion and a photon, i.e., QED.   If there were additional fields in the theory, then the high energy behavior of the dispersion would certainly be modified.  For instance, if the Lagrangian contained an interaction term like $\phi \bar \psi' \psi$ which coupled the charged fermion $\psi$ to some other charged massive fermion $\psi'$ and scalar $\phi$, then photon scattering mediated by the new fermion and scalar would contribute terms dispersive terms of order $\mathcal{O}(\omega^2)$.  For these processes, the arrival time formula in Eq.~(\ref{cosmictlag}) is valid as the inelastic threshold of this process is set by the mass of the particles $\phi$ and $\psi'$; these computations will mirror the results for the neutral  dark matter considered below.    Alternatively, we could imagine that our charged fermion $\psi$ is a composite, rather than point-like, particle;  its composite nature would also impact the particle's polarizability.  

These results can be applied to standard model particles as well.   The baryonic energy density relative to the critical density  in the universe today is $\Omega_b = 0.04628 \pm 0.00093$ \cite{wmap9}, roughly 15\% of the entire matter in the universe, and the overwhelming majority of this matter consists of atomic hydrogen.   In the long wavelength limit,  electromagnetic radiation cannot resolve the individual charged constituents of a neutral atom, so neutral hydrogen will not appreciably contribute to the refractive index for charged particles in the low photon energy limit.  In the high energy limit, well beyond the ionization energy, neutral hydrogen can be considered as a plasma of protons and electrons.  From Eqs.~(\ref{vg_correction})  and (\ref{hi_E_charged}), we see that the less massive electrons will contribute more to dispersion than protons.  But, as before, for reasonable photon energies, the correction to the $\mathcal{O}(\alpha)$ term in the refractive index is of little consequence.

Since the protons are composite particles, their interactions with photons are richer.  Indeed, the parity-violating terms, Eq.~(\ref{hslash_lot}), in the forward Compton amplitude can result in an $\mathcal{O}(\omega)$ term in the index of refraction.  Using the helicity basis for the photon polarization Eq.~(\ref{pviol_amp_circ}), we can write the $P$-violating contribution to the proton index of refraction as
\begin{equation}
n\Pslashsup_\pm(\omega) = 1 + \frac{ \rho_b}{4m_p^2 \omega^2} \left[g\Pslashsup (\omega)\,  \boldsymbol{\mathcal{S}} \cdot  \hat{\mathbf{n}}
\pm h\Pslashsup (\omega) \right].
\end{equation}
If the proton medium is unpolarized then the spin-dependent terms will average to zero $\langle \boldsymbol{\mathcal{S}} \cdot  \hat{\mathbf{n}} \rangle  =0$.  From the low-energy expansion in Eq.~(\ref{hslash_lot}), we thus have
\begin{equation}
n\Pslashsup_\pm(\omega) \approx 1 \pm \frac{ \rho_b}{4m_p^2} \frac{e^2 g_A h^{(1)}_{\pi NN} \mu_n}{ 6\sqrt{2} \pi^2 F_\pi  m_\pi^2}\, \omega.
\end{equation}
From Ref.~\cite{chen}, we find the values $g_A = 1.26$, $F_\pi = 93$ MeV, $\mu_n = -1.91$, $h^{(1)}_{\pi NN} \sim 5 \times 10^{-7}$.  For photon energies below the pion threshold, $\omega \ll 140$ MeV, we have 
\begin{equation}
n\Pslashsup_\pm(\omega) \approx 1\mp 8.8 \times 10^{-59} \frac{\omega}{\text{MeV}}.
\end{equation}
As a point of comparison, we note that relic neutrinos can render the cosmos optically active, though on a vastly different scale.  For relic neutrinos at a temperature of 2 K, the difference between the index of refraction for positive and negative helicity photons is $n_+ - n_-=  2.2 \times 10^{-78}\, \omega/\text{MeV} $ \cite{abb_rep1}.

At sufficiently high energy, the $\mathcal{O}(\omega)$ term in the index of refraction should dominate
the $A_0$ term, present for charged particles, in Eq.~(\ref{n_expans}).  For the explicit example considered here, the $\mathcal{O}(\omega)$ term for the proton is only valid below  the pion inelastic threshold.  For photon energies below this threshold, the $A_0$ term dominates; as such, the explicit terms linear in $\omega$ cannot be observed via the time-lag procedure.   Considering a broader class of particles, we do expect that parity-violating terms in the forward Compton amplitude can result in circularly birefringent medium arising from a dispersive term in the index of refraction that is linear in $\omega$.  For instance, the interaction between a photon and the electron's anapole moment violates parity \cite{e_anapole}, and this should result in such a term for photons with energy below the weak scale.

\subsection{Scalar neutral dark matter \label{boehm_sect}}

As an example of neutral dark matter, we consider the $N=2$ SUSY-inspired model of light (MeV) scalar dark matter  proposed initially in Ref.~\cite{boehm1} by Boehm and Fayet and expounded upon in Ref.~\cite{boehm2}.  
This model  was introduced as a possible explanation for excessive  511 keV radiation, seen by the INTEGRAL satellite, originating from our galactic center \cite{integral}.   
A comprehensive review of this topic can be found in Ref.~\cite{prantzos}. 
Regardless of the motivation for this model, we choose to study it in the present context because a medium consisting of scalar particles is optically simple, producing no birefringent effects.  Additionally, this model requires only a few extra particles beyond the SM, so it is a natural starting point before considering the panoply of  particles in the MSSM.  

The relevant new particles contributing to the index of refraction are the scalar dark-matter particle $\phi_\text{dm}$ and a new massive fermion $\psi_F$, both of which couple to a SM fermion such as an electron or quark $\psi_f$.  The interaction term in the Lagrangian is 
\begin{equation}
\mathcal{L}_\text{int} =  \overline{\psi}_F (g + ih\gamma^5) \psi_f \phi_\text{dm} + \text{ h.c.} \label{boehmL}
\end{equation}
In Ref.~\cite{boehm2}, DM annihilation into two photons is considered via an effective four-point interaction by integrating out the heavy fermion $F$
\begin{equation}
\mathcal{L}_\text{eff} = \frac{1}{M_F} \phi_\text{dm}^* \phi_\text{dm}  \overline{\psi}_f (a + ib\gamma^5) \psi_f.
\end{equation} 
where $M_F$ is the mass of the heavy fermion $F$.
Using this effective theory to explore the crossed process in which the photon forward scatters on the DM is fruitless as the amplitude is identically zero.  
This is not surprising when one considers the low energy theorem for the forward Compton amplitude. 
For a neutral candidate, we expect the leading order contribution to be $\mathcal{O}(\omega^2)$.  As the DM and photon four-momenta, $p$ and $q$, respectively, are the only external vectors involved in the scattering process, terms which yield factors of $\omega$ must come from the Lorentz-invariant product $p \cdot q$.   If the four point interaction in $\mathcal{L}_\text{eff}$ is used for the forward scattering process, then momentum flow through the electron loop does not involve $p$ (at least, not in a non-trivial manner).  
As a result, the effective theory can only render the $\mathcal{O}(\omega^0)$ term, which vanishes for a neutral scatterer.

\begin{figure}[th]
\includegraphics[width=3in]{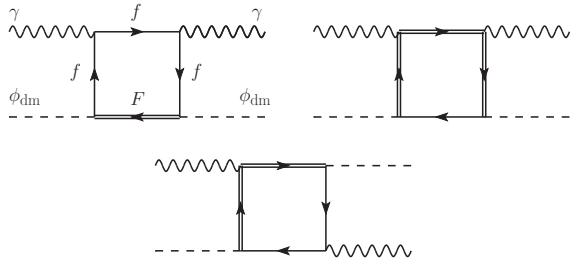}
\caption{Photon-DM scattering diagrams for the Lagrangian in Eq.~(\ref{boehmL}).  The dashed lines represent the scalar DM; the single fermion line represents the SM particle; and the double fermion line represents $F$.} \label{fig3}
\end{figure}

To go beyond this approximation, we need to consider the full  theory at the one-loop level, Eq.~(\ref{boehmL}).  This involves calculating six one-loop Feynman diagrams, including both circulations of the fermion loops. Three of the diagrams are pictured in Fig.~(\ref{fig3}) with clockwise circulation; the remaining graphs are found by crossing the photon lines.  
Each closed fermion loop results in a trace over a product of fermion propagators and vertex operators, and we use the usual Dirac trace technology to simplify these.
Simple power counting shows that each individual diagram is logarithmically divergent.  As such, we use dimensional regularization to calculate its contribution to the amplitude.   
Upon summing all terms, the result is finite.  Each graph yields a four-point integral over loop momentum; however, in the forward limit, we can utilize various tricks, such as partial fraction decomposition, to express the integrals in terms of only three Feynman parameters.     

We work in the rest frame of the DM particle with $p=(m_\text{dm}, \mathbf{0})$ and $q=(\omega, \omega  \mathbf{\hat n})$ the four-momenta of the DM and photon, respectively; we let $m_f$  be the mass of the SM fermion and $M_F$ the mass of the new heavy fermion.  The amplitude can be naturally grouped into terms which are either proportional to $g^* g - h^* h$ or $g^* g + h^*h$. We will decouple these  accordingly in the amplitude $\mathcal{M} = \mathcal{M} _- +\mathcal{M}_+$.  
We note that terms proportional to $g^*h + h^* g$  vanish identically for each diagram, and terms proportional to $g^*h - h^* g$ vanish when both orientations of fermion flow are considered. 
Performing all but the last parameter integral, we find the following contributions to the scattering amplitude
\begin{eqnarray}
\mathcal{M}_- &=& - 8 e^2 (g^* g - h^* h)    \frac{1}{(4\pi)^2}    \int_0^1\mathrm{d}x  (2 M_F m_f) \Bigg[ \frac{-1}{P_1(x)} + \frac{s}{2(p \cdot q)^2} \log \Bigg( \frac{S_1(x)}{P_1(x)} \Bigg) + \frac{u}{2(p \cdot q)^2} \log \Bigg( \frac{U_1(x)}{P_1(x)} \Bigg) \nonumber \\
&&+\frac{M_F^2 -s + m_f^2}{4(p \cdot q)^2}\Bigg( \frac{1}{x} \log\Bigg( \frac{S_1(x)}{P_1(x)} \Bigg) + \frac{1}{x}  \log\Bigg( \frac{S_2(x)}{P_2(x)} \Bigg) \Bigg) 
 + \frac{M_F^2 -u + m_f^2}{4(p \cdot q)^2}\Bigg( \frac{1}{x} \log\Bigg( \frac{U_1(x)}{P_1(x)} \Bigg) + \frac{1}{x}  \log \Bigg( \frac{U_2(x)}{P_2(x)} \Bigg)\Bigg)  \Bigg], \nonumber \\ \label{ampa} \\ 
\mathcal{M}_+ &=& -8 e^2 (g^* g +h^* h) \frac{1}{(4\pi)^2}\Bigg\{  \int_0^1\mathrm{d}x  \frac{1}{2} \Bigg[   \frac{1}{x}\log \Bigg(  \frac{S_1(x)}{P_1(x)} \Bigg) +\frac{1}{x}\log \Bigg(  \frac{U_1(x)}{P_1(x)} \Bigg) +\frac{1}{x}\log \Bigg(  \frac{S_2(x)}{P_2(x)} \Bigg) +\frac{1}{x}\log \Bigg(  \frac{U_2(x)}{P_2(x)} \Bigg)  \Bigg] \nonumber \\
&&+ [M_F^2 -p^2 +m_f^2]  \int_0^1 \mathrm{d}x \Bigg[ \frac{-1}{P_1(x)} + \frac{s}{2(p \cdot q)^2} \log \Bigg( \frac{S_1(x)}{P_1(x)} \Bigg) + \frac{u}{2(p \cdot q)^2} \log \Bigg( \frac{U_1(x)}{P_1(x)} \Bigg) \nonumber \\
&&+\frac{M_F^2 -s + m_f^2}{4(p \cdot q)^2}\Bigg( \frac{1}{x} \log\Bigg( \frac{S_1(x)}{P_1(x)} \Bigg) + \frac{1}{x}  \log\Bigg( \frac{S_2(x)}{P_2(x)} \Bigg) \Bigg) 
 + \frac{M_F^2 -u + m_f^2}{4(p \cdot q)^2}\Bigg( \frac{1}{x} \log\Bigg( \frac{U_1(x)}{P_1(x)} \Bigg) + \frac{1}{x}  \log \Bigg( \frac{U_2(x)}{P_2(x)} \Bigg)\Bigg)  \Bigg]\Bigg\}. \nonumber \\ \label{ampc}
\end{eqnarray}
We define the polynomials
\begin{eqnarray}
P_1(x) &=& p^2 (x^2-x) +(M_F^2  -m_f^2)x +m_f^2 ,\\
S_1(x) &=& s (x^2-x) +(M_F^2 -m_f^2)x +m_f^2 ,\\
U_1(x) &=& u (x^2-x) +(M_F^2 -m_f^2)x +m_f^2,
\end{eqnarray}
where the usual Mandelstam variables are $s = (p+q)^2$ and $u=(p-q)^2$.  The polynomials $P_2$, $S_2$, and $U_2$ are defined in the same manner after swapping the masses of the charged fermions $M_F \leftrightarrow m_f$.  Closed form solutions to the integrals are discussed in the Appendix.

Central to the low energy behavior of this amplitude is the claim that the leading order behavior is $\mathcal{O}(\omega^2)$.  To prove this we shall focus upon $\mathcal{M}_-$; the result for $\mathcal{M}_+$  follows rather trivially from this.
In Eq.~(\ref{ampa}), there are terms with prefactors of $\omega^{-1}$ and $\omega^{-2}$, and one term completely independent of $\omega$.  
After expanding the integrand as a power series in $\omega$, we find that the $\omega^{-1}$ and $\omega^{-2}$ terms vanish but a nonzero $\mathcal{O}(\omega^0)$ term remains
\begin{equation}
\mathcal{M}_- = -\frac{8 e^2  }{(4\pi)^2}(g^* g - h^* h) [2M_F m_f] \int_0^1 \mathrm{d}x \Bigg\{ \frac{4x^2 -4x +1}{P_1(x)}- \frac{x(1-x)[2p^2x(1-x) +(M_F^2 -p^2 +m_f^2)]}{[P_1(x)]^2} \Bigg\} + \mathcal{O}(\omega^2).\label{loint}
\end{equation}
However, the integral in Eq.~(\ref{loint}) is exactly zero as the integrand is the derivative of a function which vanishes on the boundary
\begin{equation}
 \mathcal{M}_- = -\frac{8 e^2 }{(4\pi)^2} (g^* g - h^* h) [2M_F m_f]   \int_0^1 \mathrm{d}x \frac{\mathrm{d}}{\mathrm{d}x}\Bigg(\frac{x-3x^2+2x^3}{P_1(x)}\Bigg)  + \mathcal{O}(\omega^2).
\end{equation}

To assess the leading order contribution to the dispersive index of refraction, we make further approximations. 
 The DM candidate of Boehm {\it et al}.~has a mass of MeV scale, and the new charged fermion $F$ has evaded detection in collider searches, such as LEP \cite{lep}, setting a lower bound on its mass $M_F \ge 84$ GeV.   
Given these masses, it is appropriate to consider the limit $m_f, m_\text{dm} \ll M_F$.  We approximate the amplitude in the limit $m_f \ll M_F$
\begin{eqnarray}
\mathcal{M}_- &\approx& -\frac{16 e^2 }{(4\pi)^2}(g^* g - h^* h) \frac{m_f}{M_F} \Bigg[ \frac{m_\text{dm}^2 \omega^2}{M_F^4} \Bigg(\frac{2}{3}\log \frac{M_F^2}{m_f^2} -2 \Bigg)+ \frac{m_\text{dm}^4 \omega^4}{M_F^8}\Bigg(\frac{16}{15}\log \frac{M_F^2}{m_f^2} -\frac{40}{9} \Bigg)\Bigg] + \mathcal{O}(\omega^6) \\
\mathcal{M}_+ &\approx& \frac{8 e^2 }{(4\pi)^2}(g^* g + h^* h)\Bigg[ \frac{m_\text{dm}^2 \omega^2}{M_F^4} \Bigg(\frac{4}{3}\log \frac{M_F^2}{m_f^2} -2 \Bigg)+ \frac{m_\text{dm}^4 \omega^4}{M_F^8}\Bigg(\frac{44}{15}\log \frac{M_F^2}{m_f^2} -\frac{92}{9} \Bigg)\Bigg] + \mathcal{O}(\omega^6) \label{ampcapprox},
\end{eqnarray}
where, for terms which are $\mathcal{O}(\omega^n)$, we only keep terms up to $\mathcal{O}(1/M_F^{2n})$.  We note that the $\mathcal{O}(\omega^2)$ term of the forward Compton amplitude, $\mathcal{M}_+$, for the BF model is similar in structure to the analogous term in photon-neutrino scattering \cite{abb_rep1}.   Photon-neutrino scattering is effected, at the leading order in perturbation theory, by one-loop diagrams involving virtual $W$ bosons and electrons.  Per Ref.~\cite{abb_rep1}, neglecting terms of order $\mathcal{O}(m_e^2/m_W^2)$, we find
\begin{equation}
\mathcal{M}^{\nu\gamma \to \nu \gamma}\approx \frac{\alpha^2 s^2}{2m_W^2\sin^2\theta_W} \left[ \frac{4}{3} \log\left( \frac{m_W^2}{m_e^2}\right) +1\right].
\end{equation}
Though the neutrino is a spin-1/2 particle, the leading order behavior of the forward Compton amplitude should exhibit a structure similar to the results from the BF model.  This is a useful check for our calculations.

\begin{figure}[t]
\includegraphics[width=3in]{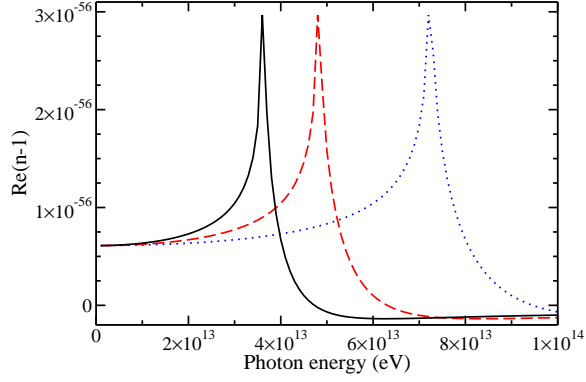}
\caption{The deviation from unity for the local index of refraction governing the phase velocity for light traveling through the scalar dark matter of Ref.~\cite{boehm2}. The solid (black) line employs $m_\text{dm} = 100$ MeV;  the dashed (red) line employs $m_\text{dm} = 75$ MeV; the dotted (blue) line employs $m_\text{dm} = 50$ MeV.  For all curves $M_F=85$ GeV and $m_f = m_e$.  The peaks in each curve signal the appearance of the inelastic threshold at this order in  perturbation theory.  At this threshold energy, the imaginary part of the index of refraction is first nonzero, and the real part exhibits anomalous dispersion beyond this energy.  
 \label{fig4}}
\end{figure}

Returning to the BF model, relative to $\mathcal{M}_+$,  the $\mathcal{M}_-$ portion of the amplitude is suppressed by an additional factor of $m_f/M_F$.  
Inserting this amplitude into the local index of refraction governing the phase speed, Eq.~(\ref{n_expans}), we see the dark matter polarizability scales as $M_F^{-4}$, and the leading dispersive term of  $\mathcal{O}(\omega^2)$ has a coefficient proportional to $M_F^{-8}$, highly suppressing dispersion.  
At the one-loop level, the index of refraction is real for $s < (M_F + m_f)^2$.  With the assumption $m_f, m_\text{dm} \ll M_F$, the photon energy needed to approach this threshold energy is $\omega \sim M_F^2/(2m_\text{dm})$.  As a point of reference, for $M_F=85$ GeV and $m_\text{dm}=100$ MeV, this photon threshold energy is around a 36 TeV.  Given this, our $\mathcal{O}(\omega^2)$ approximation of the refractive index is quite good for the photon energies accessible by the Fermi telescope \cite{fermi_lat}.
From the full expression for the amplitude, Eq.~(\ref{ampc}), we plot in Fig.~\ref{fig4} the deviation of the real part of the index of refraction from unity for various values of DM mass with  $M_F=85$ GeV and $m_f = m_e$.  To be definite, we set $g^* g + h^* h=1$ and $g^* g - h^* h=0$.  We employ the usual value for the average dark matter density $\rho_0\simeq 1.20\times 10^{-6}$ GeV/cm${}^3$. The curves in Fig.~\ref{fig4} exhibit typical behavior for the index of refraction around a resonance.  The cusps in the curves occur at the inelastic threshold at this order in perturbation theory.  Above this threshold energy, the imaginary part of the refractive index becomes nonzero.  Below the threshold energy, the real part of the refractive index increases with photon energy, and beyond the threshold the medium exhibits anomalous dispersion in which the real part of the refractive index decreases as the photon energy increases.

Though the travel time over cosmological scales will be enhanced due to density and blue-shift factors, the timing resolution needed to detect these time lags is prohibitive.  As a best case scenario, we calculate the time difference between the arrival of a low energy and 1 TeV photon for a source at the distant redshift $z=8$.   We use the group velocity with the approximate $\mathcal{O}(\omega^4)$ amplitude in Eq.~(\ref{ampcapprox}) and the parameters indicated above with $m_\text{dm}=100$ MeV.  Namely, we calculate
\begin{equation}
\tau = \frac{3 \rho_0 \omega^2}{4 m_\text{dm}^2}A_4 K_4 (z=8) ,  \label{lag_w2}
\end{equation}
where
\begin{equation}
A_4 =  \frac{2 \alpha }{\pi}(g^* g + h^* h)\frac{m_\text{dm}^4}{M_F^8}\Bigg(\frac{44}{15}\log \frac{M_F^2}{m_f^2} -\frac{92}{9} \Bigg).
\end{equation}
We note that $K_4(z=8)=3.52 \times 10^{7}$ Mpc and $A_4 = 1.03 \times 10^{-20}$ GeV$^{-4}$; the resulting time lag is prohibitively small, $\tau = 3 \times 10^{-38}$ s.

As is apparent in Fig.~\ref{fig4} and Eq.~(\ref{ampcapprox}), the scale of the dispersive effects is set by the polarizability term which is, for the present limits, independent of $m_\text{dm}$ and controlled primarily by a term of the form $M_F^{-4}$.  There is an additional logarithmic enhancement of the polarizability dependent upon the ratio $M_F/m_f$; however, this enhancement is relatively small.

\subsection{Neutral fermionic dark matter \label{nulino_sect}}

In a supersymmetric theory, if R-parity is conserved, then the lightest supersymmetric particle (LSP) is a natural dark matter candidate since it is stable to decay.  A popular candidate for the LSP is the lightest neutralino, a Majorana fermion.  A neutralino $\tilde \chi^0_j$ is a linear combination of the four neutral gauginos:  the bino $\tilde{B}^0$, wino $\tilde{W}^0$, and higgsinos $\tilde{H}^0_{1,2}$, the superpartners to the SM neutral bosons.  In fact, the neutralinos are  the mass eigenstates of these gauginos/higgsinos.  
Using the notation of Ref.~\cite{edsjo}, the matrix, $Z_{jk}$, relating the mass eigenstates to the gauginos/higgsinos is defined by
\begin{equation}
\tilde \chi^0_j = Z_{j1} \tilde{B}^0 + Z_{j2} \tilde{W}^0 + Z_{j3} \tilde{H}_1^0 + Z_{j4} \tilde{H}_2^0.
\label{nulino}
\end{equation}
The lightest neutralino corresponds to $j=1$; however, we shall drop the subscript and denote the lightest neutralino as $\tilde{\chi}^0$.  Since this dark matter candidate is  spin-1/2, this admits the possibility of a birefringent medium. Recalling the requirements for birefringence \cite{nieves_pal},  both $P$ and $CP$ asymmetry must be present in the photon-neutralino interaction and/or in the neutralino medium itself.    Since Majorana fermions are their own antiparticles, the only way for the neutralino medium to be $P$ or $CP$ asymmetric is for the medium to be polarized.  If we assume an unpolarized medium, then the conditions necessary for birefringence must be satisfied by the photon-neutralino interaction; however, in the forward scattering limit, the process must be $T$ (or $CP$) symmetric.  
In what follows, we will only consider an unpolarized medium so that the refractive index will have the form of Eq.~(\ref{n_expans}).

A full one-loop calculation of the amplitude for neutralino annihilation into two photons has been reported in Refs.~\cite{bergstrom,bern}.  The amplitude in these papers is computed in the rest frame of neutralinos so that we cannot simply make use of crossing symmetry to determine the forward Compton amplitude.  Here, we will compute, in steps, the full one-loop forward Compton amplitude in the neutralino rest frame.  First, we will  consider the contribution to the amplitude for processes involving virtual fermion and sfermions in the loop, and we include, in Appendix B, the remaining contributions which involve virtual $W^\pm$ bosons, charginos, and Higgs particles.  
Since the neutralino is a Majorana particle, the usual Feynman rules for Dirac fermions must be adapted; we follow the procedure outlined in Refs.~\cite{denner1,denner2} by Denner {\it et al.}.  Alternate rules for dealing with Majorana fermions exist, e.g., Ref.~\cite{haber_kane}, but either set of rules will yield equivalent results.  Evaluating an individual diagram will result in a scalar product which depends upon the initial and final spinor states of the neutralino.  Since we are considering completely coherent forward scattering, evaluation of an individual diagram can take the form $\boldsymbol{\epsilon}_\mu^* \boldsymbol{\epsilon}_\lambda \overline{u}^s(p) \Gamma^{\mu\lambda} u^s(p)$, for positive-energy states $u^s(p)$ of spin $s$ with $\Gamma^{\mu\lambda}$ some product of Dirac gamma functions.
Since we are considering an unpolarized neutralino medium, we would like to average over the possible spin states.  By averaging over spin states, we can turn our scalar products into traces
\begin{equation}
\frac{1}{2} \sum_{s=1,2} \overline{u}^s(p) \Gamma^{\mu\lambda} u^s(p) = \frac{1}{2} \mathrm{Tr}[\Gamma^{\mu\lambda} (\pslash + m)] ,
\end{equation}
where we have used the completeness relation $\sum_{s=1,2} u^s(p) \overline{u}^s(p) = (\pslash + m)$.
At this point, we can employ the usual Dirac trace technology to simplify the traces.

\begin{center}
{\bf Fermion-sfermion contribution}  
\end{center}

\begin{figure} [h]
 \includegraphics[width=3in]{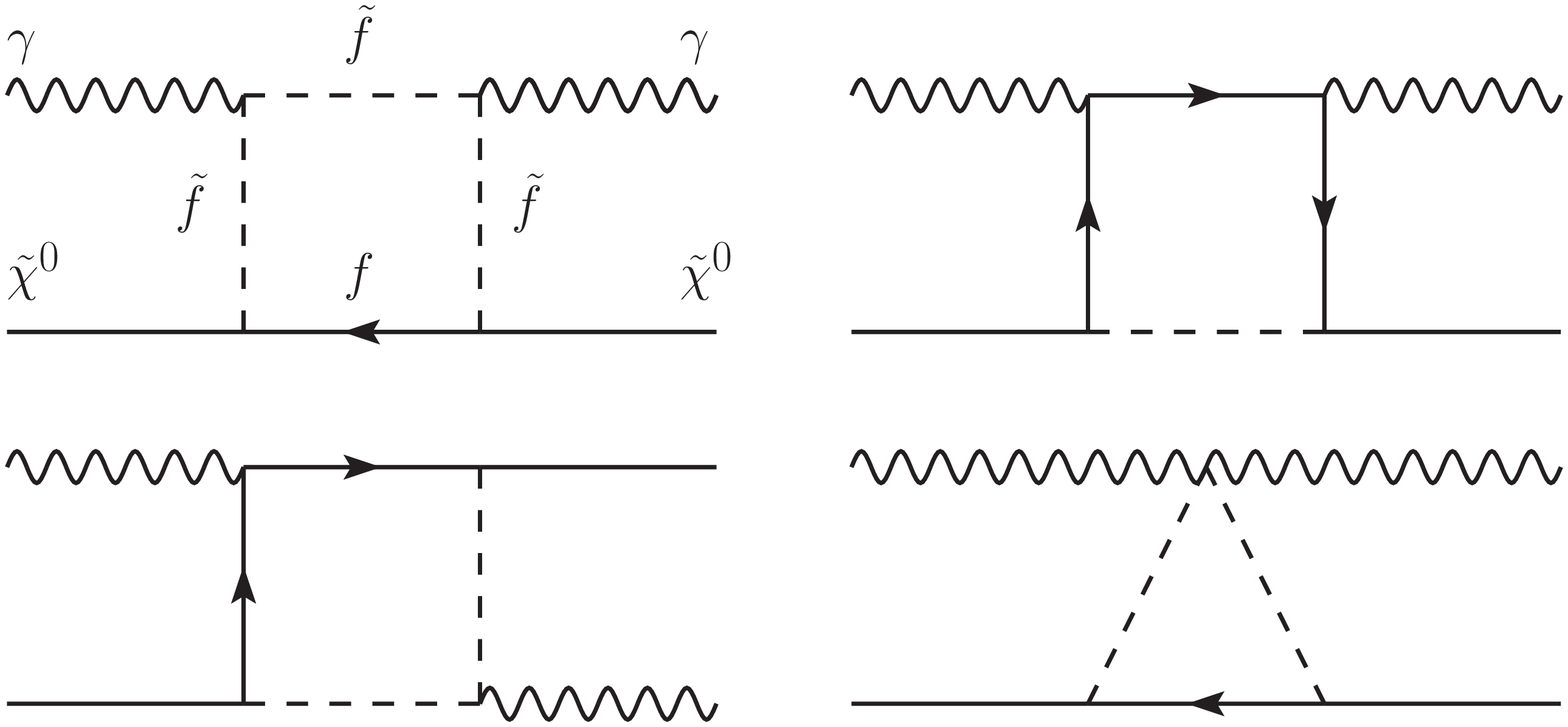} 
 \caption{Feynman graphs contributing to the forward Compton amplitude on a neutralino.  The neutralino is a Majorana fermion, which we represent  with a solid fermion line without an arrow.  The other lines follow the notation in Fig.~\ref{fig3}.  Namely, the dashed lines represent scalar particles (sfermions); the solid lines with arrows represent Dirac fermions; and the wavy lines represent spin-1 bosons (photons).
  \label{fig5}}
 \end{figure}

Some of the Feynman diagrams for  neutralino-photon forward scattering which involve virtual fermion and sfermions are shown in Fig.~\ref{fig5}.
 In addition to the graphs pictured, we need to consider both orientations of the virtual fermion line $f$ and the graphs with the ``crossed" photons. 
We adopt the notation of Ref.~\cite{edsjo} for the coupling constants.  In particular, the neutralino-sfermion-fermion vertex is expressed as $i(g^L_{\tilde{f} f 1} P_L + g^R_{\tilde{f} f 1} P_R)$, where the charged fermion's momentum flows away from the vertex and $P_{L/R} = \frac{1}{2}(1 \mp \gamma^5)$.  

As before, we represent the photon four-momentum as $q$ with $q^2=0$, and the neutralino  momentum is $p$ with $p^2 = m_{\tilde\chi^0}^2$.   We denote the fermion mass as $m_f$ and the sfermion mass as $M_{\tilde{f}}$; the charge of this fermion is $e_f e$.  The contribution from the fermion-sfermion graphs to the forward Compton amplitude is 
\begin{eqnarray}
 \mathcal{M}_{\tilde{f}f} &=& -  \frac{2 e_f^2 e^2}{(4 \pi)^2}  
\int_0^1 \mathrm{d}x  (g^{R*}_{\tilde{f} f1} g^R_{\tilde{f} f1} + g^{L*}_{\tilde{f} f 1} g^L_{\tilde{f} f 1})
\Bigg\{  
\frac{1}{2x} \log\Bigg( \frac{S_1(x)}{P_1(x)}  \Bigg)  +\frac{1}{2x} \log\Bigg( \frac{U_1(x)}{P_1(x)}\Bigg) 
   \nonumber \\
  &&  + (M_{\tilde{f}}^2-m_{\tilde\chi^0}^2 -m_f^2 ) \Bigg\lgroup - \Bigg[ \frac{1}{S_1(x)} + \frac{1}{U_1(x)} + \frac{1}{P_1(x)}   \Bigg] \nonumber \\
  &&+ \frac{1}{p\cdot q}  \Bigg[     [  M_{\tilde{f}}^2- m_{\tilde\chi^0}^2+ m_f^2] \Bigg( \frac{1}{S_1(x)} -\frac{1}{U_1(x)} \Bigg)  - \frac{1}{2x} \Bigg(  \log\Bigg( \frac{S_1(x)}{U_1(x)}  \Bigg)  +\log\Bigg( \frac{S_2(x)}{U_2(x)}  \Bigg)  \Bigg)  \Bigg]  \nonumber\\ 
 &&- \frac{1}{(p\cdot q)^2}  \Bigg[ \frac{1}{4} [M_{\tilde{f}}^2 - (m_f +m_{\tilde\chi^0})^2 ][M_{\tilde{f}}^2 - (m_f -m_{\tilde\chi^0})^2 ] \Bigg( \frac{1}{S_1(x)} - \frac{2}{P_1(x)} + \frac{1}{U_1(x)}\Bigg) \nonumber\\
 && -[M_{\tilde{f}}^2 -  m_{\tilde\chi^0}^2 +m_f^2]   \frac{1}{4x} \Bigg( \log\Bigg( \frac{S_1(x)}{P_1(x)}  \Bigg) +\log\Bigg( \frac{U_1(x)}{P_1(x)}  \Bigg)+\log\Bigg( \frac{S_2(x)}{P_2(x)}  \Bigg) +\log\Bigg( \frac{U_2(x)}{P_2(x)}  \Bigg)  \Bigg) \Bigg]     \Bigg \rgroup \Bigg\}
\nonumber\\
&&-  \frac{2 e_f^2 e^2}{(4 \pi)^2}  
\int_0^1 \mathrm{d}x (g^{R*}_{\tilde{f} f1} g^L_{\tilde{f} f1} + g^{L*}_{\tilde{f} f1} g^R_{\tilde{f} f1})2 m_{\tilde\chi^0} m_f \Bigg\{  \Bigg[ \frac{1}{S_1(x)} + \frac{1}{U_1(x)} + \frac{1}{P_1(x)}   \Bigg]     \nonumber \\&& - \frac{1}{p\cdot q}  \Bigg[  [ M_{\tilde{f}}^2 -m_{\tilde\chi^0}^2 + m_f^2 ]\Bigg( \frac{1}{S_1(x)} - \frac{1}{U_1(x)}  \Bigg) 
-\frac{1}{2x}  \Bigg( \log \Bigg( \frac{S_1(x)}{U_1(x)}\Bigg) +  \log \Bigg( \frac{S_2(x)}{U_2(x)}  \Bigg) \Bigg)  \Bigg]
\nonumber \\
&& + \frac{1}{(p\cdot q)^2}  \Bigg[ 
\frac{1}{4} [M_{\tilde{f}}^2-  (m_f +m_{\tilde\chi^0})^2 ][M_{\tilde{f}}^2-  (m_f - m_{\tilde\chi^0})^2 ] 
\Bigg( \frac{1}{S_1(x)} -\frac{2}{P_1(x)} +\frac{1}{U_1(x)} \Bigg)    \nonumber \\
&& -[M_{\tilde{f}}^2 -m_{\tilde\chi^0}^2 + m_f^2 ]   \frac{1}{4x} \Bigg( \log\Bigg( \frac{S_1(x)}{P_1(x)}  \Bigg) +\log\Bigg( \frac{U_1(x)}{P_1(x)}  \Bigg)+\log\Bigg( \frac{S_2(x)}{P_2(x)}  \Bigg) +\log\Bigg( \frac{U_2(x)}{P_2(x)}  \Bigg)  \Bigg) \Bigg]  
\Bigg\}. \label{ampff}
 \end{eqnarray}
We adapt the notation from the previous section and set
  \begin{eqnarray}
     P_1(x)&=&  p^2(x^2-x) +(M_{\tilde{f}}^2 - m_f^2)x +m_f^2 ,\\
   S_1(x)&=& s(x^2-x) +(M_{\tilde{f}}^2- m_f^2)x + m_f^2 ,\\
     U_1(x)&=& u(x^2-x) +(M_{\tilde{f}}^2 - m_f^2)x + m_f^2 , \label{polynomials_ff}
  \end{eqnarray}
  with $s=(p+q)^2 = m_{\tilde\chi^0}^2+ 2 p \cdot q$ and $u=(p-q)^2 = m_{\tilde\chi^0}^2 -2 p \cdot q$.  The polynomials $P_2(x)$, $S_2(x)$, and $U_2(x)$ are defined as above under the interchange $M_{\tilde{f}} \leftrightarrow m_f$.
Using the symbolic manipulation program FORM \cite{form}, the leading order term in an expansion about $\omega$ is shown to be $\mathcal{O}(\omega^2)$; we find

\begin{eqnarray}
\mathcal{M}_{\tilde{f} f} &=&   \frac{2 e_f^2 e^2}{(4 \pi)^2}  
(q \cdot p)^2 \int_0^1 \mathrm{d}x  \Bigg\{ 
(g^{R*}_{\tilde{f} f} g^R_{\tilde{f} f1} + g^{L*}_{\tilde{f} f1} g^L_{\tilde{f} f1})
\Bigg[   \frac{2x(1-x)^2 }{(P_1(x))^2} \Bigg( 1+ (q \cdot p)^2 \frac{2x^2(1-x)^2 }{(P_1(x))^2}  \Bigg) \nonumber \\
&&  -\frac{2}{3}(M_{\tilde{f}}^2-m_{\tilde\chi^0}^2-m_f^2) \frac{x^2(1-x)^2}{(P_1(x))^3} \Bigg( 1 +
(q \cdot p)^2  \frac{8}{5} \frac{x^2(1-x)^2}{(P_1(x))^2} \Bigg)
 \Bigg] \nonumber \\
&&
+(g^{R*}_{\tilde{f} f1} g^L_{\tilde{f} f1} + g^{L*}_{\tilde{f} f1} g^R_{\tilde{f} f1}) m_{\tilde\chi^0} m_f
 \frac{4}{3} \frac{x^2(1-x)^2 }{(P_1(x))^3} \Bigg[1  +(q \cdot p)^2  
 \frac{8}{5} \frac{x^2(1-x)^2 }{(P_1(x))^2}  \Bigg] \Bigg\} + \mathcal{O}(\omega^6). \label{nulinopol}
\end{eqnarray}

\begin{center}
{\bf Numerics} 
\end{center}

The MSSM contains well over a hundred parameters.  Models which attempt to describe a mechanism  of symmetry breaking at the GUT scale introduce additional assumptions which pare down the list of free parameters within the particular model; however, more general considerations can still produce viable SUSY models.  
In Refs.~\cite{susy_predj, susy_predj_dm,pMSSM_nulino_gtino,pMSSM_fine,pMSSM_more}, the authors explore p(henomenological)MSSM,
searching a large swath of SUSY parameter space with no regard to the method of symmetry breaking.  Their searches employ 19 real parameters subject to collider and cosmological constraints as well as some basic theoretical constraints.  The models are CP-conserving with minimal flavor violation; the first two generations of sfermions are degenerate; and the LSP is a conventional thermal relic.  Germane to our present work are the masses of the supersymmetric particles in viable models.

 In comparing  Eq.~(\ref{nulinopol}) with Eq.~(\ref{wloop_lot}) in Appendix B, we see that the leading order contributions to the forward Compton amplitude coming from quark-squark, $W$-chargino, and Higgs-chargino box diagrams have similar functional forms.  Recalling the formula for the index of refraction, Eq.~(\ref{qft_n}), we see that the scale at which the index deviates from unity is set by the masses of virtual particles in Figs.~\ref{fig5} and \ref{fig8}.   In particular, the dominant mass particle in the loop sets the scale; this fact was also borne out in the BF dark matter model in Sec.~\ref{boehm_sect}.   Generally speaking, lighter masses of particles in the loop will result in a greater contribution to the refractive index.   Given this, we will focus only upon the contributions coming from the fermion-sfermion loops since the SM quarks (aside from the top) and leptons have masses much smaller than the $W$ boson, Higgs, and any SUSY charged particles.  
 
 \begin{figure}[t]
\includegraphics[width=3in]{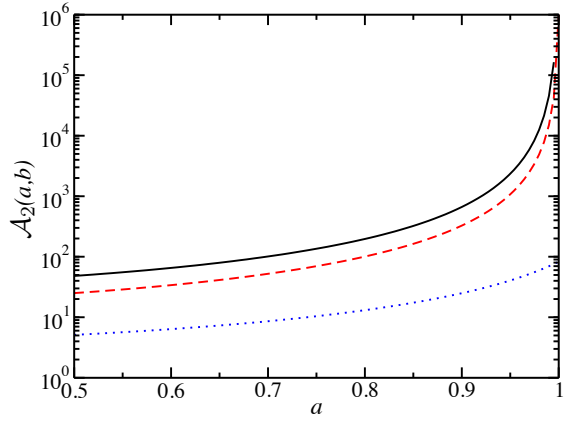}
\caption{We plot $\mathcal{A}_2(a,b)$ versus $a=m_{\tilde\chi^0}/M_{\tilde{f}}$.  The solid (black) curve corresponds to $b=m_f/ M_{\tilde{f}}=10^{-5}$; the dashed (red) curve has $b=10^{-3}$; and the dotted (blue) curve has $b=10^{-1}$. 
\label{fig6}}
\end{figure}
Focusing upon the fermion-sfermion loops, we make some approximations as a means to understand which SUSY parameters might maximize the size of dispersive effects in Eq.~(\ref{nulinopol}).  As most of the SM fermions entering the loop will have masses at a relatively lower scale, the term proportional to $(g^{R*}_{\tilde{f} f1} g^R_{\tilde{f} f1} + g^{L*}_{\tilde{f} f1} g^L_{\tilde{f} f1})$ in Eq.~(\ref{nulinopol}) should dominate; we analyze this piece as a function of the input masses. 
To do so, we define the dimensionless function $\mathcal{A}_2$ as
\begin{equation}
\frac{1}{M_{\tilde{f}}^4} \mathcal{A}_2(a,b) =  \int_0^1 \mathrm{d}x 
\Bigg[  \frac{2x(1-x)^2 }{(P_1(x))^2}
-\frac{2}{3}(M_{\tilde{f}}^2-m_{\tilde\chi^0}^2-m_f^2) \frac{x^2(1-x)^2}{(P_1(x))^3} \Bigg]
\end{equation}
with the ratio of masses $a=m_{\tilde\chi^0}/M_{\tilde{f}}$ and $b=m_f/M_{\tilde{f}}$. 
The actual dispersive part of the index of refraction is governed by the terms which are $\mathcal{O}(\omega^4)$ in the amplitude; we define the dimensionless $\mathcal{A}_4$ to be
\begin{equation}
\frac{1}{M_{\tilde{f}}^8} \mathcal{A}_4(a,b) =  \int_0^1 \mathrm{d}x 
\Bigg[  \frac{4x^3(1-x)^4 }{(P_1(x))^4}
-\frac{16}{15}(M_{\tilde{f}}^2-m_{\tilde\chi^0}^2-m_f^2) \frac{x^4(1-x)^4}{(P_1(x))^5} \Bigg].
\end{equation}
In the limit that $b\to 0$, i.e., $m_f \to 0$, one may evaluate the integrals rather easily
\begin{equation}
 \mathcal{A}_2(a,b) =  \frac{4}{3(1-a^2)^2}\log\Bigg(\frac{(1-a^2)^2}{b^2} \Bigg) +\frac{2}{a^4}\log \Bigg(\frac{1}{1-a^2}\Bigg) - \frac{2}{a^2(1-a^2)^2}
\end{equation} 
and
\begin{equation}
 \mathcal{A}_4(a,b) =  \frac{44}{15(1-a^2)^4}\log\Bigg(\frac{(1-a^2)^2}{b^2} \Bigg) +\frac{4}{a^8}\log \Bigg(\frac{1}{1-a^2}\Bigg) + \frac{1}{(1-a^2)^4} \Bigg(-\frac{4}{a^6}+\frac{14}{a^4} -\frac{52}{3a^2} -\frac{26}{9} \Bigg).
\end{equation} 
Indeed, it is now clear that a small value of $m_f$ relative to $M_{\tilde{f}}$ can lead to a slight logarithmic enhancement of the dispersive term.

\begin{figure}[t]
\includegraphics[width=3in]{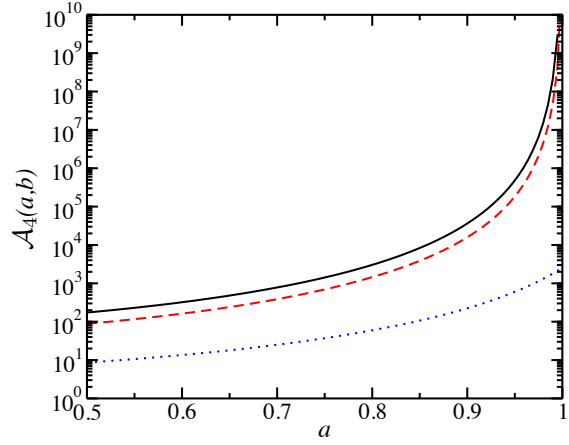}
\caption{We plot $\mathcal{A}_4(a,b)$ versus $a=m_{\tilde\chi^0}/M_{\tilde{f}}$.  The solid (black) curve corresponds to $b=m_f/ M_{\tilde{f}}=10^{-5}$; the dashed (red) curve has $b=10^{-3}$; and the dotted (blue) curve has $b=10^{-1}$. 
 \label{fig7}}
\end{figure}
In Figs.~\ref{fig6} and \ref{fig7},  we plot $\mathcal{A}_2(a,b)$ and $\mathcal{A}_4(a,b)$ for various values of $b$ as a function of $a$.  We only plot values of $a$ up to unity since we assume the lightest neutralino to be the LSP.  
In general, we see that both functions are largest whenever $m_{\tilde\chi^0} \sim M_{\tilde{f}}$ and $m_f \ll M_{\tilde{f}}$.  Given the prefactor of $M_{\tilde{f}}^{-4}$ and $M_{\tilde{f}}^{-8}$, we also note that the polarizability and dispersive effects will be largest when $M_{\tilde{f}}$ is smallest.
In Refs.~\cite{pMSSM_fine,pMSSM_more}, the scan of SUSY parameter space admits models which have third generation squarks with masses on the order of hundreds of GeV.  Additionally, a large fraction of these models have a nearly pure wino LSP.  To estimate the size of dispersion from a viable neutralino dark matter candidate, we will assume a pure wino LSP of mass 100 GeV and a sbottom of mass 200 GeV.  For a pure wino, we have in Eq.~(\ref{nulino}) $Z_{12}=1$ with $Z_{1k}=0$ otherwise; as a result, the only nonzero sbottom-bottom-wino coupling is $g_{\tilde{b}_L b 1}^R= \frac{1}{\sqrt{2}}g$ with $g$ the usual weak coupling constant.  With this notation, the leading order deviation of the local refractive index from unity for these fermion-sfermion loops is
\begin{equation}
n_{\tilde{b}b}-1 =      \frac{  \alpha^2}{18 \sin^2\theta_W} \frac{\rho_0}{M_{\tilde{b}}^4}  
\Bigg(  \mathcal{A}_2(a,b) +  \frac{m_{\tilde\chi^0}^2 \omega^2}{M_{\tilde{b}}^4} \mathcal{A}_4(a,b) \Bigg)
+\mathcal{O}(\omega^4). \label{pol_chi}
\end{equation} 
where the arguments of $\mathcal{A}_j$ are $a=0.5$ and $b=0.02$ so that $\mathcal{A}_2\sim 12$ and $\mathcal{A}_4\sim 33$.  
In terms of the time lag accrued between simultaneously emitted high (energy $\omega$) and low energy photons from a distant GRB, we have
\begin{equation}
\tau =  \frac{  \alpha^2}{6 \sin^2\theta_W} \frac{\rho_0 m_{\tilde{\chi}_0}^2}{M_{\tilde{b}}^8}  \omega^2 \mathcal{A}_4(a,b) K_4 (z).
\end{equation}
As a best case scenario for detecting such a time lag, we assume a high energy photon $\omega=100$ GeV can be detected from a very distant GRB at $z=8$; this yields a time lag $\tau \approx 1.6 \times 10^{-39}$ s.

\section{Observational considerations}

Gamma ray bursts are a useful tool by which we may realize constraints on the coefficients $A_j$ in the index of refraction expansion, Eq.~(\ref{n_expans}), because they are extremely bright emissions of high energy photons of a rather short duration.   In order to assess reliable limits on the refractive index from the study of photon arrival times, we must separate propagation effects from intrinsic source effects. 
On average, we expect time delays intrinsic to the source to be independent of $z$, and 
the time delay from propagation to depend on $z$ and $\omega$  as in Eq.~(\ref{cosmictlag}). Thus, from a large sample of GRBs occurring at various redshifts $z$, the photon energy and redshift dependence of time lags due to photon-matter interactions should be discernible from intrinsic time delays associated with GRBs.  Such notions have been previously employed in searches for Lorentz invariance violation (LIV)~\cite{ellis}.  
In this case, frequency dependent time lags for high energy photons are also expected due to a modified dispersion relation, and observations from distant GRBs {\em do} provide meaningful constraints on the energy scale at which Lorentz invariance violation (LIV) can occur.  Following Ref.~\cite{jacobpiran}, LIV effects can modify the dispersion relation for photons at an energy scale $E_\text{LIV}$
\begin{equation}
E^2 - p^2 = \pm p^2 \left( \frac{p}{E_\text{LIV}} \right)^n \label{disp_rel}
\end{equation}
for some integer $n\ge 1$.  If the modification in the dispersion relation in Eq.~(\ref{disp_rel_g}) comes with the minus sign (rather than the plus sign), then high energy photons will lag simultaneously emitted lower energy photons.  For a (present day) photon energy $\omega$ emitted at redshift $z$, this time lag is approximately
\begin{equation}
\Delta t_\text{LIV} = \frac{1+n}{2H_0} \left(\frac{\omega}{E_\text{LIV}}\right)^n \int_0^z \frac{(1+z')^n\mathrm{d}z'}{\sqrt{\Omega_M(1+z')^3 + \Omega_\Lambda}}.  \label{t_liv}
\end{equation}

To place constraints on a cosmic refractive index, we can look to the recent observations of high energy photons by the Fermi telescope \cite{fermi_sci, fermi_nat}.  In Ref.~\cite{fermi_sci}, several multi-GeV photons were detected from GRB 080916C which is located at a redshift $z= 4.35 \pm 0.15$ 
\cite{080916C_red}.  The highest energy photon, with energy $\omega_\text{high} = 13.22^{+0.70}_{-1.54}$ GeV, arrived 16.54 s after initial $\mathcal{O}(\text{MeV})$ photons from the GRB.  If this time lag were attributable to interactions with dark matter, we can assess the magnitude of the coefficient $A_4$ in the index of refraction expansion.  We recall Eq.~(\ref {lag_w2}) which expresses the time lag engendered by a high energy photon in a neutral dark-matter medium
\begin{equation}
\tau = \frac{3 \rho_0 \omega_\text{high}^2}{4 m^2}A_4 K_4 (z);
\end{equation}
as before, we assume the $\omega_\text{low}^2$ term for the MeV photons is relatively negligible. 
From the redshift of GRB 080916C, we calculate $K_4(z=4.35) = 3.33 \times 10^6$ Mpc.  The limits on $A_4$ and the dark-matter mass $m$ are thus 
\begin{equation}
\frac{A_4}{m^2} < 4.00 \times 10^{25} \text{ GeV}^{-6}.
\end{equation}

An even higher energy photon was observed from GRB 090510 \cite{fermi_nat}.  The redshift of this GRB was measured to be $z=0.903\pm 0.003$ .  The highest energy photon detected is $\omega_\text{high}=30.53^{+5.79}_{-2.56}$ GeV some 0.829 s after the trigger of the GRB monitor.  In extracting their limits on Lorentz invariance, the authors of Ref.~\cite{fermi_nat} conservatively use the 1-$\sigma$ lower bound on the photon energy, $\omega_\text{high}=28$ GeV, a time lag of 0.859 s, and the 1-$\sigma$ lower bound on the redshift, $z=0.900$.  Using these values, we find the limit
\begin{equation}
\frac{A_4}{m^2} < 6.51  \times 10^{25} \text{ GeV}^{-6}.  
\end{equation}
Despite a higher energy photon and smaller time lag, this limit is slightly weaker than that achieved from GRB 080916C.  This is due to the fact that GRB 080916C occurred at a much higher redshift.
In the context of the 
MSSM neutralino calculation from Sect.~\ref{nulino_sect}, these experimental limits do not result in any meaningful constraints on the LSP.

It is interesting to explore under what circumstances dispersive effects due to dark matter might confound measurements designed to ascertain LIV. We see that for $n=2$ the LIV time lag depends on energy as $\omega^2$, the same dependence as for neutral dark-matter dispersive effects.  For both the BF and MSSM models, the index of refraction has a similar structure.  If the charged particles in the loop are dominated by a particle of mass $M$, then we can approximate the dispersive term in the forward scattering amplitude via $A_4 \approx \alpha g^2 m^4/M^8$ for a DM candidate mass $m$. 
This structure follows from an EFT analysis of the forward Compton amplitude for a neutral particle. For photon energies below threshold, the amplitude can be approximated as $\mathcal{M}_{\lambda \lambda} \approx A_2 \omega^2 + A_4\omega^4$ in the scatterer's rest frame.  In this frame, the only way that $\omega$ can make an appearance in the amplitude is through the Lorentz invariant quantity $p \cdot k = m \omega$.  Since the amplitude is itself dimensionless, we find balancing factors of $M$ in the denominator, since this mass dominates the loop process.  A scaling argument then indicates $A_2 \omega^2 \sim m^2 \omega^2/M^4$ and $A_4 \omega^4 \sim m^4 \omega^4/M^8$.  Combining this with the index of refraction formula, Eq.~(\ref{qft_n}), we see that the $1/m^2$ prefactor which multiplies the amplitude is cancelled by the factor of $m^2$ common to all the amplitude terms.  Hence, with this naive analysis, low-mass scatterers will not necessarily boost the size of the dispersive terms.

Taking the dispersive term as $A_4 \approx \alpha g^2 m^4/M^8$, the  time lag is then approximately
\begin{equation}
\tau \approx  \rho_0 \frac{ \alpha g^2m^2 }{M^8} \frac{\omega^2}{H_0} \int_0^z  \frac{(1+z')^{5}\mathrm{d}z'}{\sqrt{\Omega_M(1+z')^3 + \Omega_\Lambda}} .\label{approx_tau}
\end{equation}  
Though the integrands in Eqs.~(\ref{t_liv},\ref{approx_tau}) carry different powers of $z'$, they only differ by factor of order $\mathcal{O}(1)$ and can be neglected in an order of magnitude calculation.  Thus, we see that dark-matter dispersive effects could confound $n=2$ LIV effects whenever
\begin{equation}
 \alpha g^2\frac{ \rho_0  m^2 }{M^8} \gtrsim E_\text{LIV}^{-2}. \label{ineq}
\end{equation}
Weakly interacting massive particles (WIMPs) are a favored DM candidate because weak-scale interactions and DM masses naturally produce the correct relic density $\Omega_{dm} \sim m_w^2/g_w^4$; this is the so-called WIMP miracle.  However, it was shown in Ref.~\cite{wimpless} that alternative DM models can still achieve the desired relic density if 
\begin{equation}
\frac{m}{g^2} \sim \frac{ m_w}{g_w^2};
\end{equation}
this is referred to as the WIMPless miracle.  Invoking this relation, we define the DM coupling in terms of the DM mass, $g^2=g^2_w m/m_w$. Inserting this into the inequality in Eq.~(\ref{ineq}), we find that DM dispersive effects are important whenever the DM mass satisfies
\begin{equation}
\frac{ M^8} {m^3} \lesssim    {\alpha g_w^2}\frac{ \rho_0 E_\text{LIV}^2}{m_w} . 
\end{equation}
Futhermore, stability of the DM candidate requires $m< M$, so that we have
\begin{equation}
 m^5 \lesssim    {\alpha g_w^2}\frac{ \rho_0 E_\text{LIV}^2}{m_w} .   \label{m_liv}
\end{equation}
To estimate the size of the relevant DM mass, let us suppose that LIV effects occur at the Planck scale $E_\text{LIV}\approx 10^{19}$ GeV.  Setting $g_w = 0.65$ and $m_w = 100$ GeV, we find that DM dispersive effects could confound LIV effects  whenever $m \lesssim 2$ MeV.  However, to saturate this inequality, the dominant mass particle, $M$, to which the DM couples must be on the same MeV scale, that is $m\sim M$.  This presents two difficulties.  First, collider constraints limit unit-charged exotic particles to have masses well beyond this, $M > 85$ GeV.  Second, if this could somehow be evaded, an MeV scale mass $M$ would limit the inelastic threshold for the photon-DM interaction to also be on the MeV scale.  This would result in a dispersive medium only below the MeV threshold with no relevant dispersive effects above this energy.   Given this, if LIV effects occur near the Planck scale, then frequency dependent time-lags that scale as $\omega^2$ cannot be confused with DM dispersive effects because the matter effects are not appreciable.   

Turning this on its head, suppose we evade collider constraints on the mass of exotic charged particles by setting the DM mass and dominant mass to be $m\sim M\sim 100$ GeV.   Then the inequality, Eq.~(\ref{m_liv}), indicates that matter effects are competitive with $n=2$ LIV effects at an LIV energy scale 
$ E_\text{LIV} \gtrsim \sqrt{m^5 m_w/(\alpha g_w^2\rho_0)} \approx 10^{29}$ GeV -- well beyond the GZK cutoff \cite{gzk}. Below this energy, an observation of time lags that correlate with photon energy as $\omega^2$ cannot be due to dispersive matter effects for the ilk of neutral DM models considered herein.

The same GRBs above can be used to constrain the LIV scale for an $n=1$ modification of the photon dispersion relation, Eq.~(\ref{disp_rel}).  GRB 080916C indicates $E_\text{LIV}> 1.3 \times 10^{18}$ GeV \cite{fermi_sci}, and GRB 090510 indicates $E_\text{LIV} > 1.22 \times 10^{19}$ GeV \cite{fermi_nat}.   Recalling from above, an $\mathcal{O}(\omega)$ term can appear in the index of refraction if there is both $P$ and $CP$ asymmetry present in the photon-matter interactions or the medium itself.  Generally, this term results in circular birefringence of the medium, but no model independent expression exists for the $\mathcal{O}(\omega)$ term in the index of refraction, making it difficult to make generic remarks about the energy scale at which matter effects could compete with those attributable to LIV.  However, one class of quantum gravity models--stringy models of spacetime foam \cite{Ellis:1999uh,ellis_foam,Ellis:2005wr,ellis2,Ellis:2009yx}--results in a modified dispersion relation which {\em can} be distinguished from matter effects.  
Using an EFT approach, one can induce LIV effects by including five- and six-dimensional operators in a QED Lagrangian  \cite{myers-pospelov}.  The resulting $\mathcal{O}(\omega)$ modifications to the photon dispersion relation will necessarily result in vacuum birefringence.  To contrast, spacetime foam models move beyond the EFT paradigm and can result in a modified photon dispersion relation which is linear in photon energy but {\em not} birefringent.  Since matter effects  will always result in a birefringent medium for  $\mathcal{O}(\omega)$ energies, photon time lags that scale linearly with photon energy {\em and} are polarization independent signal LIV due to a QG effect like spacetime foam.

\section{Conclusion}

From some rather basic assumptions, we determine the index of refraction for light traveling through particulate matter.  In previous work, we investigated the optical consequences of a DM particle with fractional electric charge \cite{millicharge}, where the leading order contribution to the refractive index scales with photon energy as $\omega^{-2}$.   Herein, we considered higher order contributions coming from the $\mathcal{O}(\alpha^2)$ corrections to the forward Compton amplitude.  The dominant term in this correction is a quasi-static polarizability which results in a term in the index of refraction that scales as $\log(\omega/m)$ at low photon energies; however, this term is dominated by the leading order term in perturbation theory.  At higher energies, likewise, the corrections are negligible for photon energies below the GZK cutoff.   

For neutral DM candidates, the leading order contribution to the forward Compton amplitude comes from the polarizability of the particle for photon energies below the inelastic threshold. As a consequence, the leading order term in the refractive index is  frequency independent, and the largest dispersive term is quadratic in photon energy.  At the one-loop level in perturbation theory, we explicitly confirmed this behavior for a neutral scalar DM candidate as well as for MSSM's neutralino.  For both models, we considered the situation in which one of the charged particles in the loop had a dominant mass $M$.  For this situation, the dispersive term in the index of refraction scaled as $ \rho_0  m^2 \omega^2 /M^8$ for DM mass $m$ and energy density $\rho_0$.  Given the collider constraints $M>85$ GeV, dispersion from even a distant GRB results in a negligible time lag between simultaneously emitted high and low energy photons.  If lighter charged exotic particles did in fact exist, say, on the MeV scale, then the cosmos could have measurable dispersion. 

Since neutral dark matter can result in time lags which scale with photon energy as $\omega^2$, we compared the size of photon time lags arising from a dispersive medium with the $\mathcal{O}(\omega^2)$ time lags that can result from Lorentz invariance violation.  For the models considered herein, in which the neutral DM candidate couples to a massive charged exotic particle, we find $\mathcal{O}(\omega^2)$ time lags due to LIV will dominate matter effects below the GZK cutoff.

\section{Appendix A}
The following integrals appear in the forward-scattering box diagrams
\begin{eqnarray}
I_1(s,m_1^2,m_2^2) &=& \int_0^1 \mathrm{d}x \frac{1}{P(s,m_1^2,m_2^2;x)}, \\
I_2(s,m_1^2,m_2^2) &=& \int_0^1 \mathrm{d}x \log [P(s,m_1^2,m_2^2;x)], \\
I_3(s,m_1^2,m_2^2) &=&  \int_0^1 \mathrm{d}x \frac{1}{x} \log \Bigg[ \frac{P(s,m_1^2,m_2^2;x)}{m_1^2}\Bigg],
\end{eqnarray}
with the polynomial defined to be $P(s,m_1^2,m_2^2;x)=(x^2-x)s +(m_2^2-m_1^2)x +m_1^2$.  Here, the masses of the virtual particles are $m_1$ and $m_2$; the four-momentum of the scatterer is $p$ with $p^2=m^2$; and the four momentum of the photon is $q$ with $q^2=0$.  The relevant Mandelstam variables are $s=(p+q)^2$ and $u=(p-q)^2$.
We define the threshold energy $E_\text{th}^2=(m_1+m_2)^2$ and mass difference $\Delta^2=(m_2-m_1)^2$.  We assume that the particle is stable to decay so that $p^2 < E_\text{th}^2$.
The first argument of the functions $I_j$ will be either $s$, $u$, or $p^2$.  

We consider  the case where $s$ is the first argument of $I_j$.
The integrals can be divided into three different regions $s< \Delta^2$, $\Delta^2 < s < E_\text{th}^2$, and $s > E_\text{th}^2$.   All three integrals are real except when $s > E_\text{th}^2$. 
We note that $I_j (s,m_1^2,m_2^2)= I_j (s,m_2^2,m_1^2)$ for $j=1,2$ (but {\em not} $j=3$).  This is true because $P(s,m_1^2,m_2^2;x)=P(s,m_2^2,m_1^2;1-x)$ so that, after a change of variables in the second integral, we have
\begin{equation}
\int_0^1 \mathrm{d}x f[P(s,m_1^2,m_2^2;x)] = \int_0^1 \mathrm{d}x f[P(s,m_2^2,m_1^2;x)]
\end{equation}
for some function $f$ whose argument is the polynomial $P$.

 \bigskip
We begin with $I_1$ and $I_2$ and discuss $I_3$ later.
For $s < \Delta^2$, we have
\begin{eqnarray}
I_1(s,m_1^2,m_2^2) &=&  \frac{2}{\sqrt{(E_\text{th}^2-s)(\Delta^2-s)}}\log\Bigg( \frac{\sqrt{E_\text{th}^2-s}+\sqrt{\Delta^2-s}}{\sqrt{E_\text{th}^2-s}-\sqrt{\Delta^2-s}} \Bigg), \\
I_2(s,m_1^2,m_2^2) &=& \log(m_1 m_2) +\frac{m_2^2 - m_1^2}{2s} \log\Bigg( \frac{m_2^2}{m_1^2} \Bigg) -2 +\frac{\sqrt{(E_\text{th}^2 - s)(\Delta^2 -s)}}{s}\log \Bigg( \frac{\sqrt{E_\text{th}^2 - s}-\sqrt{\Delta^2 - s}}{\sqrt{E_\text{th}^2 - s}+\sqrt{\Delta^2 - s}} \Bigg).
\end{eqnarray}

For $\Delta^2 < s < E_\text{th}^2$, we have
\begin{eqnarray}
I_1(s,m_1^2,m_2^2) &=&  \frac{2}{\sqrt{(E_\text{th}^2-s)(s-\Delta^2)}}\Bigg[ \tan^{-1} \Bigg( \frac{m_2^2-m_1^2+s }{\sqrt{(E_\text{th}^2-s)(s-\Delta^2)}} \Bigg) -  \tan^{-1} \Bigg( \frac{m_2^2-m_1^2- s }{\sqrt{(E_\text{th}^2-s)(s-\Delta^2)}} \Bigg) \Bigg] ,\\
I_2(s,m_1^2,m_2^2) &=&  \log(m_1 m_2) +\frac{m_2^2 - m_1^2}{2s} \log\Bigg( \frac{m_2^2}{m_1^2} \Bigg) -2  \nonumber \\
&&+\frac{\sqrt{(E_\text{th}^2 - s)(s-\Delta^2)}}{s} \Bigg[ \tan^{-1} \Bigg( \frac{m_2^2-m_1^2+s }{\sqrt{(E_\text{th}^2-s)(s-\Delta^2)}} \Bigg) -  \tan^{-1} \Bigg( \frac{m_2^2-m_1^2- s }{\sqrt{(E_\text{th}^2-s)(s-\Delta^2)}} \Bigg) \Bigg] .
\end{eqnarray}

For $s > E_\text{th}^2$, we have
\begin{eqnarray}
I_1(s,m_1^2,m_2^2) &=& \frac{2}{\sqrt{(s-E_\text{th}^2)(s-\Delta^2)}}\log\Bigg( \frac{\sqrt{s-\Delta^2}-\sqrt{s-E_\text{th}^2}}{\sqrt{s-\Delta^2}+\sqrt{s-E_\text{th}^2}} \Bigg), \\
I_2(s,m_1^2,m_2^2) &=& \log(m_1 m_2) +\frac{m_2^2 - m_1^2}{2s} \log\Bigg( \frac{m_2^2}{m_1^2} \Bigg) -2  +\frac{\sqrt{(s-E_\text{th}^2)(s-\Delta^2)}}{s}\log \Bigg( \frac{\sqrt{s-\Delta^2}+\sqrt{s- E_\text{th}^2}}{\sqrt{s-\Delta^2}-\sqrt{s- E_\text{th}^2}} \Bigg).
\end{eqnarray}

To deal with $I_3$, we first factor the polynomial in the following manner
\begin{equation}
P(s,m_1^2,m_2^2;x) = m_1^2 (A_1^+x + 1)(A_1^-x +1)  \label{factor}
\end{equation} 
with 
\begin{equation}
A_1^\pm = \frac{1}{2m_1^2} \left( m_2^2-m_1^2-s \pm \sqrt{(s-\Delta^2)(s-E_\text{th}^2)} \right).
\end{equation}
With this decomposition, the integral can be evaluated in terms of dilogarithms
\begin{equation}
I_3(s,m_1^2,m_2^2) =  -\mathrm{Li}_2(-A_1^+) - \mathrm{Li}_2(-A_1^-).
\end{equation}

In certain situations, the $I_3$ integrals come in pairs as the sum $I_3(s,m_1^2,m_2^2)+I_3(s,m_2^2,m_1^2)$.
This second integral can be evaluated in exactly the same way
\begin{equation}
I_3(s,m_2^2,m_1^2) = -\mathrm{Li}_2(-A_2^+) - \mathrm{Li}_2(-A_2^-)
\end{equation}
with the analogous definition
\begin{equation}
A_2^\pm = \frac{1}{2m_2^2} \left( m_1^2-m_2^2-s \pm \sqrt{(s-\Delta^2)(s-E_\text{th}^2)} \right).
\end{equation}
With these integrals in pairs, it turns out that the dilogarithms in the amplitude may be completely eliminated via the identity from Ref.~\cite{lewin}
\begin{equation}
\mathrm{Li}_2(-x) + \mathrm{Li}_2\Big(\frac{x}{x+1}\Big) = -\frac{1}{2} \log^2(x+1),
\end{equation}
valid for $x > -1$.    
This expression is operative as 
\begin{equation}
-A_{1}^\pm = \frac{A_{2}^\mp}{A_{2}^\mp +1}, \quad -A_{2}^\pm = \frac{A_{1}^\mp}{A_{1}^\mp +1} 
\label{id}.
\end{equation}
To prove this, we recall the relation $P(s,m_1^2,m_2^2; x) = P(s,m_2^2,m_1^2; 1-x)$. Employing the factorization from Eq.~(\ref{factor}), one has
\begin{equation}
m_1^2 (A_1^+x + 1)(A_1^-x +1) = m_2^2 (A_2^+(1-x) + 1)(A_2^-(1-x) +1).
\end{equation}
Each of the two roots of the polynomial on the LHS corresponds to one of the roots of the polynomial on the RHS yielding the identities in Eq.~(\ref{id}).

As before, we express the real part of this equation for the three different intervals of $s$.
For $s< \Delta^2$, we have
\begin{eqnarray}
I_3(s,m_1^2,m_2^2)+I_3(s,m_2^2,m_1^2) &=& \frac{1}{2}\Bigg[ \log^2\Bigg( \frac{m_1^2+m_2^2-s + \sqrt{(\Delta^2-s)(E_\text{th}^2-s)} }{2m_1^2} \Bigg) \nonumber \\
&&+ \log^2\Bigg( \frac{m_1^2+m_2^2 -s + \sqrt{(\Delta^2-s)(E_\text{th}^2-s)} }{2m_2^2} \Bigg)\Bigg].
\end{eqnarray}
For $\Delta^2 < s < E_\text{th}^2$, we have 
\begin{equation}
I_3(s,m_1^2,m_2^2)+I_3(s,m_2^2,m_1^2) =  \log^2\Bigg( \frac{m_2}{m_1} \Bigg) - 
\Bigg[ \tan^{-1}\Bigg( \frac{\sqrt{(s-\Delta^2)(E_\text{th}^2-s)} }{m_1^2+m_2^2-s} \Bigg) + \pi \Theta(s-m_1^2 - m_2^2) \Bigg]^2
\end{equation}
with $\Theta(x)$ the Heaviside function.
Finally, when $s \ge E_\text{th}^2$, we have
\begin{eqnarray}
I_3(s,m_1^2,m_2^2)+I_3(s,m_2^2,m_1^2) &=&-\pi^2 +\frac{1}{2}\Bigg[ \log^2\Bigg( \frac{s- m_1^2-m_2^2 - \sqrt{(s- \Delta^2)(s-E_\text{th}^2)} }{2m_1^2} \Bigg) \nonumber \\
&&+ \log^2\Bigg( \frac{s- m_1^2-m_2^2- \sqrt{(s-\Delta^2)(s-E_\text{th}^2)} }{2m_2^2} \Bigg)\Bigg].
\end{eqnarray}

When evaluating integrals of the form $I_j(p^2,m_1^2,m_2^2)$ and $I_j(u,m_1^2,m_2^2)$, one only needs to substitute $p^2$ or $u$ for $s$ in the above forumlae.  We note that if the dark matter particle is stable to decay then $p^2 < E_\text{th}^2$ so that one need not consider the intervals where $u,p^2 > E_\text{th}^2$ as this is not physically possible.

\section{Appendix B}

In this Appendix, we include the remaining contributions to the forward Compton amplitude of the neutralino.  These calculations  involve $W$-chargino and Higgs-chargino loops.  

\begin{center}
{\bf $W$-chargino contribution} 
\end{center}

As discussed in Ref.~\cite{bergstrom}, opting to work in a non-linear $R_\xi$ gauge greatly simplifies the calculations.  Proposed by Fujikawa \cite{fujikawa}, this choice of gauge eliminates vertices between photons, $W$ bosons, and the unphysical Higgs bosons $G$, and it simplifies the $W$ boson propagators.   A detailed list of the Feynman rules in this framework can be found in Ref.~\cite{gavela}; in our calculations, we set $\xi=1$.

\begin{figure} [h]
 \includegraphics[width=3in]{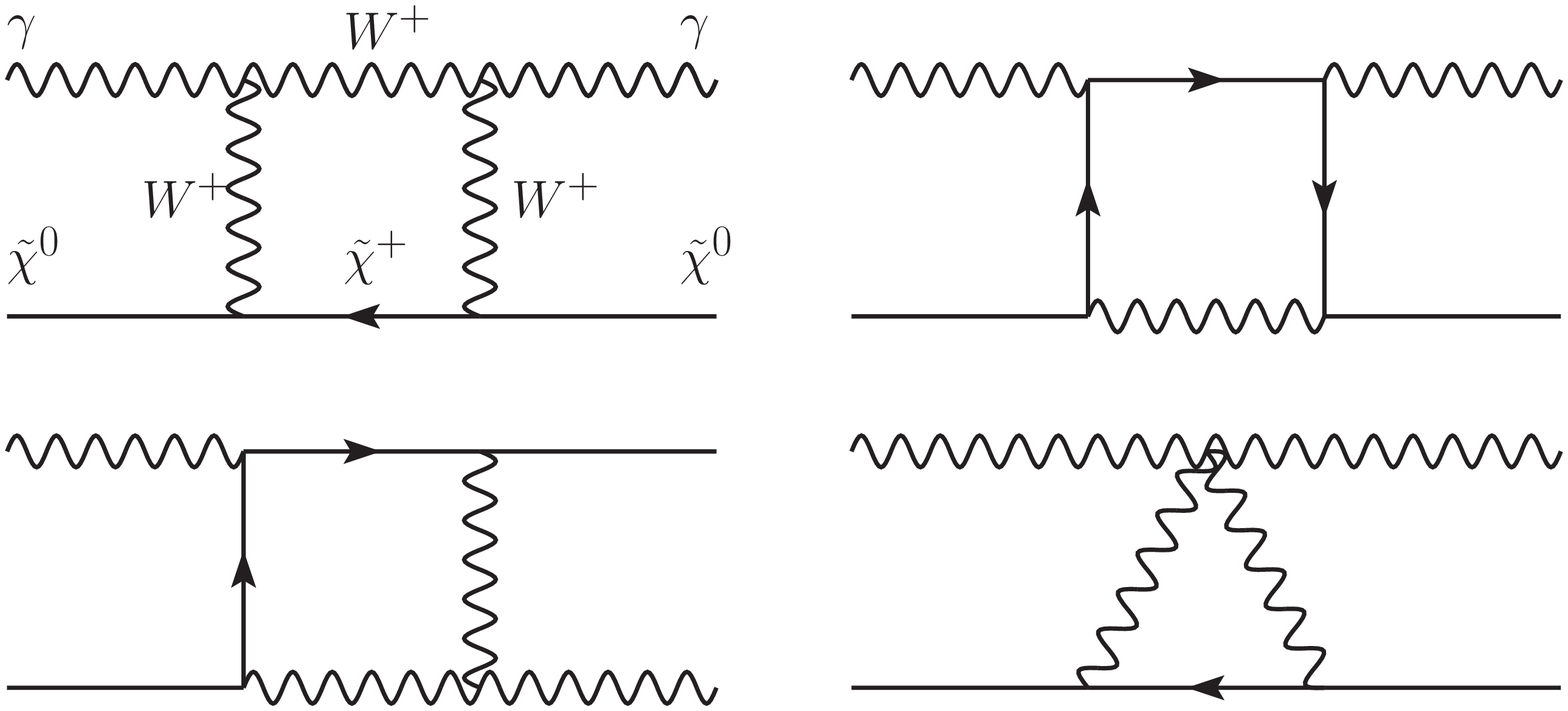} 
 \caption{Feynman diagrams involving the $W$ boson and charginos contributing to the forward Compton amplitude for a neutralino.
  \label{fig8}}
 \end{figure}

Figure \ref{fig8} contains the Feynman diagrams needed to compute the contribution to the forward Comptom amplitude coming from the $W$-chargino loops.  There are four such charginos $\tilde\chi^+_j$ indexed by $j$, and the coupling between it, the $W$ boson, and the lightest neutralino is denoted by $g^{L,R}_{W1j}$, depending on whether we project onto the left- or right-handed state.

The forward scattering amplitude for these processes follows 

\begin{eqnarray}
 \mathcal{M}_{W\tilde\chi^+} &=& -  \frac{2 e^2}{(4 \pi)^2}  
\int_0^1 \mathrm{d}x  (g^{R*}_{W1j} g^R_{W1j} + g^{L*}_{W1j} g^L_{W1j})
\Bigg\{  
\frac{1}{x} \log\Bigg( \frac{S_1(x)}{P_1(x)}  \Bigg)  +\frac{1}{x} \log\Bigg( \frac{U_1(x)}{P_1(x)}\Bigg) 
   \nonumber \\
  &&  +  (M_W^2-m_{\tilde\chi^0}^2 -m_{\tilde\chi^+}^2 ) \Bigg\lgroup - 2 \Bigg[ \frac{1}{S_1(x)} + \frac{1}{U_1(x)} + \frac{1}{P_1(x)}   \Bigg] \nonumber \\
  &&+ \frac{1}{p\cdot q}  \Bigg[    2 [  M_W^2- m_{\tilde\chi^0}^2+ m_{\tilde\chi^+}^2] \Bigg( \frac{1}{S_1(x)} -\frac{1}{U_1(x)} \Bigg)  - \frac{1}{x} \Bigg(  \log\Bigg( \frac{S_1(x)}{U_1(x)}  \Bigg)  +\log\Bigg( \frac{S_2(x)}{U_2(x)}  \Bigg)  \Bigg)  \Bigg]  \nonumber\\ 
 &&- \frac{1}{(p\cdot q)^2}  \Bigg[ \frac{1}{2} [M_W^2 - (m_{\tilde\chi^+} +m_{\tilde\chi^0})^2 ][M_W^2 - (m_{\tilde\chi^+} -m_{\tilde\chi^0})^2 ] \Bigg( \frac{1}{S_1(x)} - \frac{2}{P_1(x)} + \frac{1}{U_1(x)}\Bigg) \nonumber\\
 && -[M_W^2 -  m_{\tilde\chi^0}^2 +m_{\tilde\chi^+}^2]   \frac{1}{2x} \Bigg( \log\Bigg( \frac{S_1(x)}{P_1(x)}  \Bigg) +\log\Bigg( \frac{U_1(x)}{P_1(x)}  \Bigg)+\log\Bigg( \frac{S_2(x)}{P_2(x)}  \Bigg) +\log\Bigg( \frac{U_2(x)}{P_2(x)}  \Bigg)  \Bigg) \Bigg]     \Bigg \rgroup 
\nonumber\\
 && -\frac{1}{M_W^2} (M_W^2-(m_{\tilde\chi^0}+m_{\tilde\chi^+})^2) (M_W^2-(m_{\tilde\chi^0}-m_{\tilde\chi^+})^2)  \Bigg[ \frac{1}{S_1(x)} -\frac{2}{P_1(x)}+\frac{1}{U_1(x)}\Bigg] \nonumber \\
&& +\frac{4 p \cdot q}{M_W^2} ( M_W^2  - m_{\tilde\chi^0}^2+m_{\tilde\chi^+}^2  )  \Bigg[ \frac{1}{S_1(x)} -\frac{1}{U_1(x)}\Bigg]   -\frac{4(p \cdot q)^2}{M_W^2} \Bigg[ \frac{1}{S_1(x)} +\frac{1}{U_1(x)}\Bigg]  \Bigg\}
\nonumber\\
&& -  \frac{2 e^2}{(4 \pi)^2}  
\int_0^1 \mathrm{d}x 
(g^{R*}_{W1j} g^L_{W1j} + g^{L*}_{W1j} g^R_{W1j})2 m_{\tilde\chi^0} m_{\tilde\chi^+}  \Bigg\{ -4 \Bigg[ \frac{1}{S_1(x)} + \frac{1}{U_1(x)} + \frac{1}{P_1(x)}   \Bigg]     \nonumber \\
&& + \frac{1}{p\cdot q}  \Bigg[  4 [ M_W^2 -m_{\tilde\chi^0}^2 + m_{\tilde\chi^+}^2 ]\Bigg( \frac{1}{S_1(x)} - \frac{1}{U_1(x)}  \Bigg) 
-\frac{2}{x}  \Bigg( \log \Bigg( \frac{S_1(x)}{U_1(x)}\Bigg) +  \log \Bigg( \frac{S_2(x)}{U_2(x)}  \Bigg) \Bigg)  \Bigg]
\nonumber \\
&& + \frac{1}{(p\cdot q)^2}  \Bigg[ - [M_W^2-  (m_{\tilde\chi^+} +m_{\tilde\chi^0})^2 ][M_W^2-  (m_{\tilde\chi^+} - m_{\tilde\chi^0})^2 ] 
\Bigg( \frac{1}{S_1(x)} -\frac{2}{P_1(x)} +\frac{1}{U_1(x)} \Bigg)    \nonumber \\
&& +[M_W^2 -m_{\tilde\chi^0}^2 + m_{\tilde\chi^+}^2 ]   \frac{1}{x} \Bigg( \log\Bigg( \frac{S_1(x)}{P_1(x)}  \Bigg) +\log\Bigg( \frac{U_1(x)}{P_1(x)}  \Bigg)+\log\Bigg( \frac{S_2(x)}{P_2(x)}  \Bigg) +\log\Bigg( \frac{U_2(x)}{P_2(x)}  \Bigg)  \Bigg) \Bigg]  
\Bigg\}.
 \end{eqnarray}
The polynomials $P_1(x)$, $S_1(x)$, and $U_1(x)$  which appear in this amplitude are defined in Eq.~(\ref{polynomials_ff}) after substituting $M_{\tilde f} \mapsto M_W$ and $m_f \mapsto m_{\tilde \chi^+}$; as before, the polynomials $P_2(x)$, $S_2(x)$, and $U_2(x)$ can be obtained from these by switching the masses $M_W \leftrightarrow m_{\tilde \chi^+}$.

Using FORM \cite{form} to expand the integrands in a power series of $\omega$, we find the leading order behavior of the amplitude is $\mathcal{O}(\omega^2)$ as expected; we compute the expansion to include the leading order dispersive effects
 \begin{eqnarray}
 \mathcal{M}_{W\tilde\chi^+}  &=&   \frac{2 e^2}{(4 \pi)^2}      
 (q\cdot p)^2 \int_0^1\mathrm{d}x \Bigg\{  (g^{R*}_{W1j} g^R_{W1j} + g^{L*}_{W1j} g^L_{W1j}) \bigg[
 \frac{4x(1+x^2)}{(P_1(x))^2} \bigg(1  + (p\cdot q)^2 \frac{2x^2 (1 - x)^2}{(P_1(x))^2}\bigg)  \nonumber \\
 &&   - (  M_W^2- m_{\tilde\chi^+}^2  -  m_{\tilde\chi^0}^2 )\frac{4}{3}\frac{x^2(x-1)^2}{(P_1(x))^3} \bigg(1 + (p\cdot q)^2 \frac{8}{5} \frac{x^2(1-x)^2}{(P_1(x))^2}  \bigg)  \bigg] \nonumber \\
&& -  (g^{R*}_{W1j} g^L_{W1j} + g^{L*}_{W1j} g^R_{W1j}) m_{\tilde \chi^0}  m_{\tilde \chi^+}  \frac{16}{3}\frac{  x^2 (   1- x)^2}{(P_1(x))^3}  \left[ 1+ (p \cdot q)^2 \frac{8}{5}
 \frac{  x^2 (   1- x)^2}{(P_1(x))^2} 
   \right]\Bigg\} +\mathcal{O}(\omega^6).  \label{wloop_lot}
\end{eqnarray}

\begin{center}
{\bf Higgs-chargino contribution} 
\end{center}

Two contributions to the forward Compton amplitude remain.  The first is due to box diagrams involving the Higgs boson $H^+$ and charginos, and the second, a consequence of our gauge choice, is due to the box diagrams involving the unphysical Higgs boson $G^+$ and charginos.  In both cases, the box diagrams are identical to those in Fig.~\ref{fig5} as long as we identify the internal scalar lines with the (un)physical Higgs bosons and the internal fermion line with the charginos.  Given this similarity, the (un)physical Higgs-chargino amplitude will take the same form as Eqs.~(\ref{ampff},\ref{nulinopol}).  For the physical Higgs-chargino diagrams, we merely need to substitute the mass $M_{\tilde f} \mapsto M_{H^+}$, $m_f \mapsto m_{\tilde \chi^+}$ and couplings $g^{L,R}_{\tilde f f1} \mapsto g^{L,R}_{H^+ 1j}$ for the $j$th chargino.   For the unphysical Higgs loop, we must substitute  $M_{\tilde f} \mapsto M_{W}$ and $g^{L,R}_{\tilde f f1} \mapsto g^{L,R}_{G1j}$; explicit expressions for these couplings can be found in Ref.~\cite{bergstrom}.

\section{ACKNOWLEDGMENTS}
The author is grateful for useful comments from Susan Gardner who helped shape the trajectory of this work.
This work is supported, in part, by the U.S. Department of Energy under contract DE--FG02--96ER40989.

\end{document}